\documentclass[a4paper,english,prl,reprint,superscriptaddress]{revtex4-1}
\usepackage[T1]{fontenc}
\usepackage[latin9]{inputenc}
\usepackage{amsmath}
\usepackage{amssymb}
\usepackage{babel}
\usepackage{graphicx}

\usepackage[rgb]{xcolor}

\usepackage{siunitx}
\usepackage[normalem]{ulem}
\usepackage{textcomp}

\definecolor{OliveGreenRGB}{rgb}{0,0.6,0}

\usepackage{hyperref}
\usepackage{natbib}

\usepackage{newfloat}
\DeclareFloatingEnvironment[name={Extended Data Figure}]{suppfigure}
\DeclareFloatingEnvironment[name={Extended Data Table}]{supptable}

\usepackage{booktabs}

\begin{document}

\title{Quantum back action evading measurement of motion\\in a negative mass reference frame}

\author{Christoffer B. M{\o}ller}
\author{Rodrigo A. Thomas}
\affiliation{Niels Bohr Institute, University of Copenhagen, DK-2100 Copenhagen,
Denmark}

\author{Georgios Vasilakis}
\affiliation{Niels Bohr Institute, University of Copenhagen, DK-2100 Copenhagen,
Denmark}
\affiliation{Institute for Electronic Structure and Laser, Foundation for Research and Technology-Hellas, Heraklion 71110, Greece}

\author{Emil Zeuthen}
\affiliation{Institute for Theoretical Physics \& Institute for Gravitational
Physics (Albert Einstein Institute), Leibniz Universit\"{a}t Hannover,
Callinstra{\ss}e 38, 30167 Hannover, Germany}
\affiliation{Niels Bohr Institute, University of Copenhagen, DK-2100 Copenhagen,
Denmark}

\author{Yeghishe Tsaturyan}
\author{Kasper Jensen}
\author{Albert Schliesser}
\affiliation{Niels Bohr Institute, University of Copenhagen, DK-2100 Copenhagen,
Denmark}

\author{Klemens Hammerer}

\affiliation{Institute for Theoretical Physics \& Institute for Gravitational
Physics (Albert Einstein Institute), Leibniz Universit\"{a}t Hannover,
Callinstra{\ss}e 38, 30167 Hannover, Germany}

\author{Eugene S. Polzik}
\thanks{To whom correspondence should be addressed. Email: polzik@nbi.ku.dk}
\affiliation{Niels Bohr Institute, University of Copenhagen, DK-2100 Copenhagen,
Denmark}

\begin{abstract}
Quantum mechanics dictates that a continuous measurement of the position of an object imposes a random back action perturbation on its momentum. This randomness translates with time into position uncertainty, thus leading to the well known uncertainty on the measurement of motion. Here we demonstrate that the quantum back action on a macroscopic mechanical oscillator measured in the reference frame of an atomic spin oscillator can be evaded. The collective quantum measurement on this novel hybrid system of two distant and disparate oscillators is performed with light.  The mechanical oscillator is a drum mode of a millimeter size dielectric membrane and the spin oscillator is an atomic ensemble in a magnetic field.  The spin oriented along the field corresponds to an energetically inverted spin population and realizes an effective negative mass oscillator, while the opposite orientation corresponds to a positive mass oscillator. The quantum back action is evaded in the negative mass setting and is enhanced in the positive mass case.  The hybrid quantum system presented here paves the road to entanglement generation and distant quantum communication between mechanical and spin systems and to sensing of force, motion and gravity beyond the standard quantum limit.
\end{abstract}

\maketitle

Continuous measurement of an oscillator position, $\hat{x}(t)=\hat{x}(0)\cos (\Omega t)+\hat{p}(0)\sin (\Omega t)/(m\Omega)$, where $\Omega$ is the frequency and $m$ is the mass, leads to accumulation of the quantum back action (QBA) of the measurement in both the position and momentum, $\hat{p}$, non-commuting variables $[\hat{x},\hat{p}]=i\hbar$ \cite{Caves80,Braginsky}. Measurement QBA was recently observed for a mechanical oscillator \cite{Regal} and for atomic motion \cite{Stamper-Kurn}. Suppose, however, that the position is measured relative to an oscillator with a mass $m_0=-m$ for which $\dot{\hat{x}}_0=-\hat{p}_0/m$. The result of a measurement of $\hat{x}(t)-\hat{x}_0(t)=(\hat{x}(0)-\hat{x}_0(0))\cos (\Omega t)+(\hat{p}(0)+\hat{p}_0(0))\sin (\Omega t)/(m\Omega)$ depends only on commuting variables, $[\hat{x}-\hat{x}_0,\hat{p}+\hat{p}_0]=0$.  Hence it can be QBA free~\cite{Polzik,Hammerer} and the uncertainty in the measurement of the relative position $\langle(\hat{x}-\hat{x}_0)^2\rangle$ can be smaller than the uncertainty  $\langle\hat{x}^2\rangle$.  The first proposal for such a measurement based on atomic spins  \cite{Hammerer}, has been followed by a number of proposals for QBA free measurements \cite{Tsang,Bariani,Meystre,Clerk}.  In \cite{Caves} the negative mass approach referred to as ``quantum-mechanics-free subsystems'' was shown to lead to a measurement sensitivity approaching the Cram\'{e}r-Rao bound. Earlier work on atomic spin ensembles utilized the negative mass property for demonstration of entanglement of macroscopic spins \cite{Julsgaard} and for entanglement-assisted magnetometry \cite{Wasilewski}. The back action evading measurement on two mechanical oscillators at room temperature was demonstrated in \cite{Heidmann} in the classical regime using light, and recently in the quantum regime at the millikelvin temperature range using microwaves \cite{Sillanpaa}. Ways to overcome QBA limitations for a free mass oscillator with squeezed light have been proposed in \cite{Unruh,Kimble,Khalili}.

\begin{figure}
\begin{centering}
\includegraphics[width=1\columnwidth]{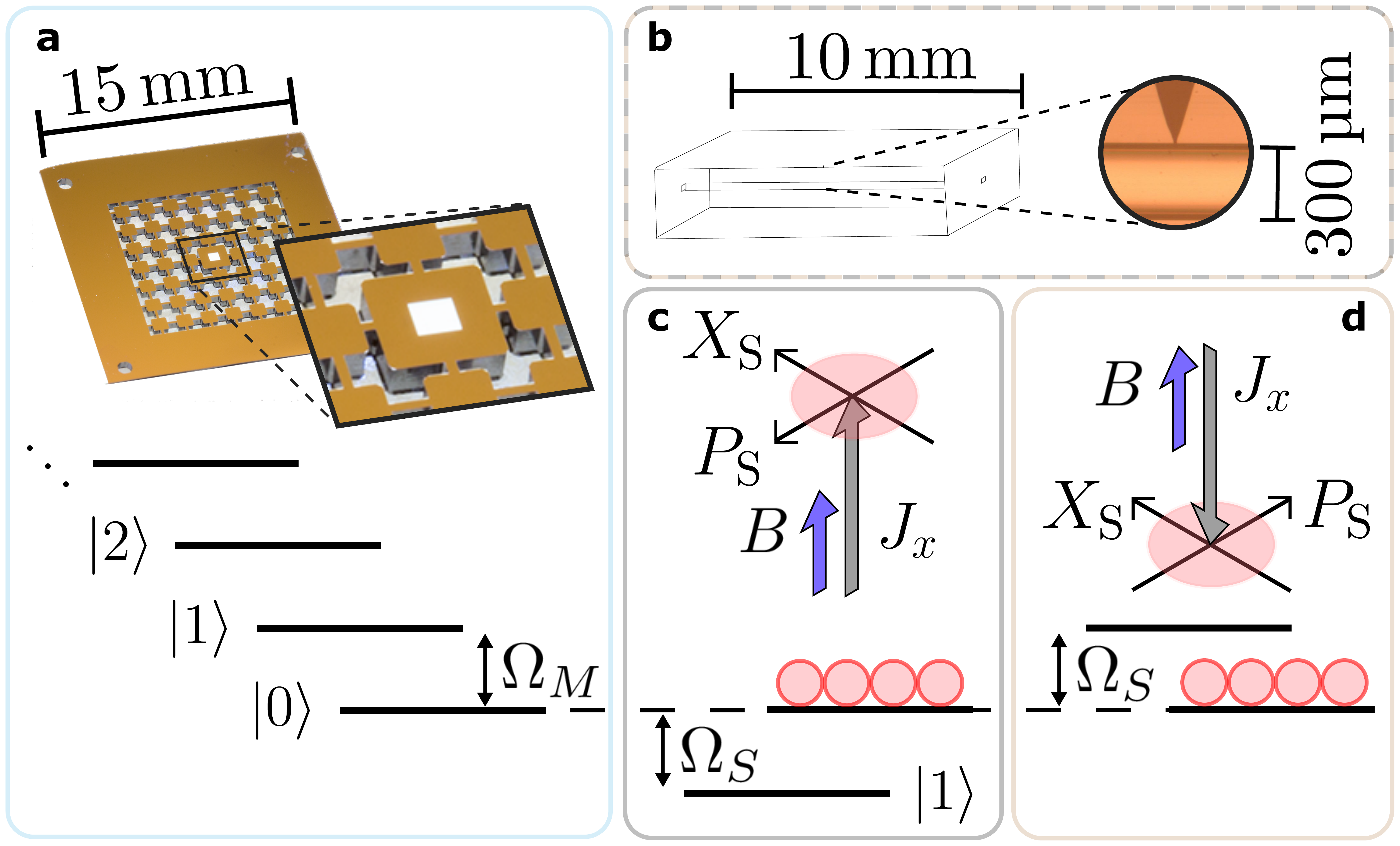}
\end{centering}
\caption{
\textbf{Mechanical and spin oscillators}. A. The mechanical oscillator -- the (1,2) drum mode, $\Omega_M=2\pi\times$\SI{1.28}{\mega\hertz}, of a \SI{0.5}{\milli\meter}, square silicon nitride membrane (light square in the center of the inset) supported by the silicon phononic crystal structure. B. The spin oscillator is an optically pumped gas of Cesium atoms contained in square crossection channel inside a glass cell. Channel walls are coated with a spin-protecting coating. The cell is placed in a static magnetic field with the Larmor frequency $\Omega_S$ tunable around  $\Omega_M$. Depending on the direction of the magnetic field with respect to the direction of the atomic spin, the oscillator can have lower (higher) energy of the excited state, corresponding to the negative (positive) effective mass, as shown in C and D, respectively.}\label{Fig:oscillators}
\end{figure}
\
Here we demonstrate QBA in a novel hybrid quantum system \cite{HybridReview,Treutlein} composed  of a macroscopic mechanical oscillator, a high-$Q$ dielectric membrane \cite{Yeghishe,Nielsen} (Fig.~\ref{Fig:oscillators}a) in a high finesse cavity, and a spin oscillator, an ensemble of room temperature Cesium atoms in a magnetic field contained in a spin-protecting environment~\cite{RMP,Seltzer2010,Vasilakis} (see Fig.~\ref{Fig:oscillators}b and Supplementary Information). The mechanical oscillator Hamiltonian is
$\hat{H}_M= (m\Omega_M^2/2)\hat{x}^2+\hat{p}^2/2m=(\hbar \Omega_M/2)(\hat{X}^2_M+\hat{P}^2_M)$,  where we henceforth employ dimensionless variables $\hat{X}_M=\hat{x}/x_{\text{zpf}}$ and $\hat{P}_M=\hat{p}\,x_{\text{zpf}}/\hbar$ where $x_{\text{zpf}}=\sqrt{\hbar/m\Omega_M}$ is the oscillator's zero point position fluctuation and $[\hat{X}_M,\hat{P}_M]=i$. Compared to a mechanical oscillator, a spin oscillator has some rather unique properties. Consider a collective atomic spin $\hat{J}_\alpha=\sum_{i=1}^{N_a}\hat{F}^i_\alpha$ with components $\alpha=x,y,z$ composed of a large number $N_a$ of ground state spins $\hat{F}^i$ (with $F=4$ in the present case). Atoms are optically pumped to generate an energetically inverted spin population in an external magnetic field $B$ (Fig.~\ref{Fig:oscillators}c), which we take to point in the positive $x$-direction. The collective spin thus exhibits a large average projection $J_x=|\langle \hat{J}_x \rangle|/\hbar \gg 1$. Its normalized $y,z$ quantum components form canonical oscillator variables $[\hat{X}_S,\hat{P}_S]=[\hat{J}_z/\sqrt{\hbar J_x },-\hat{J}_y/\sqrt{\hbar J_x}]=i$ \cite{RMP} in terms of which the spin Hamiltonian becomes $\hat{H}_{S}= \hbar\Omega_S\hat{J}_x =\hbar\Omega_S J_x -(\hbar\Omega_{S}/2)(\hat{X}_{S}^2+\hat{P}_{S}^2)$ with $\Omega_S$ -- the Larmor frequency. The first term is an irrelevant constant energy offset due to the mean spin polarization. The second term is equivalent to the Hamiltonian of a mechanical oscillator $\hat{H}_M$ with a \textit{negative} mass. Each quantum of excitation in the negative mass spin oscillator physically corresponds to a deexcitation of the inverted spin population by $\hbar\Omega_S$ (Fig.~\ref{Fig:oscillators}c). Preparation of the collective spin in the energetically lowest Zeeman state realizes instead a \textit{positive} mass spin oscillator with $\hat{H}_{S} =-\hbar\Omega_S J_x +(\hbar\Omega_{S}/2)(\hat{X}_{S}^2+\hat{P}_{S}^2)$ (Fig.~\ref{Fig:oscillators}d).

\begin{figure*}
\center\includegraphics[width=1.6\columnwidth]{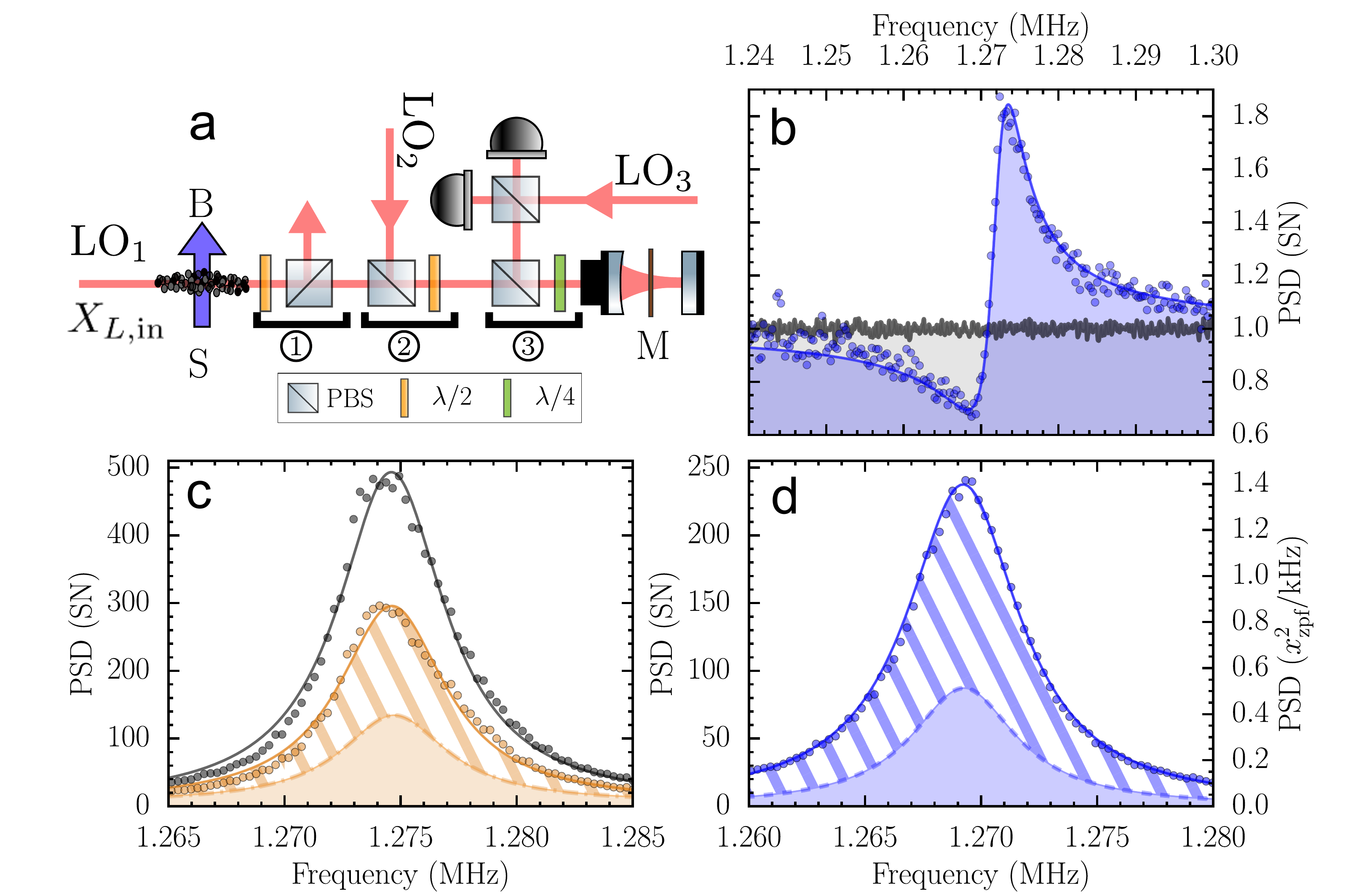}
\caption{\textbf{Experimental setup and observation of QBA for the spin and mechanical oscillators}. A. Atomic spin ensemble $S$ in magnetic field $B$ is probed by the field LO$_1$. The quadrature $X_{L,\text{in}}$ in the polarization mode orthogonal to the mean polarization of LO$_1$ is the back action (BA) force. The LO$_1$ is filtered out with the first polarizing beam splitter (PBS$_1$), while transmitted quantum fluctuations are superimposed with the field LO$_2$ at PBS$_2$, projected into the same polarization as LO$_2$ at PBS$_3$, and become a driving force for the mechanical oscillator $M$. PBS$_3$ and the quarter wave plate ensure that almost all light reflected off the cavity is directed to the homodyne detection with LO$_3$.  B. Amplitude noise spectrum of the optomechanical system showing frequency dependent squeezing of light. C. Phase noise spin spectrum. Black dots -- spin driven with the broadband thermal light noise and thermal force, brown dots -- spin driven by vacuum light noise and thermal force, brown area -- thermal noise of the spin. Striped area -- quantum back action determined from the data (see Supplementary Information). D. Phase noise of optomechanical system driven by vacuum light noise and thermal force. Blue area -- membrane thermal noise. Striped area -- quantum back action determined from squeezing data shown in A) (see Supplementary Information).  Axes labels: (SN) -- shot noise of light, $x_{\text{zpf}}$ -- zero point fluctuations. Curves are generated by the detailed numerical model of the experiment (Supplementary Information). See comments in the text.}
\label{Fig:setup}
\end{figure*}

The experiment implementing a quantum measurement on the hybrid system is sketched in Fig.~\ref{Fig:setup}a, which depicts the cascaded interaction between a traveling light field and the two oscillators (see Supplementary Information for details).
A coherent optical field with a strong, classical, linearly polarized component LO$_1$ (photon flux $\Phi_1$) and vacuum quantum fluctuations in the polarization orthogonal to it, described by quadrature phase operators $\hat{X}_{L\text{,in}}$ and $\hat{P}_{L\text{,in}}$, first interacts with the spin oscillator.
The interaction for far-off-resonant light is of the quantum nondemolition (QND) type \mbox{$\hat{H}_{\text{int},S}\propto  \hat{X}_{S}\hat{X}_{L\text{,in}}$}, where $\hat{X}_{S} \propto \hat{J}_{z}$ is the projection of the collective spin on the direction of light propagation \cite{RMP}. The light output quadrature, $\hat{P}_{L\text{,out}}^{S}(\Omega) =  \hat{P}_{L\text{,in}}(\Omega)+\sqrt{\Gamma_{S}}\hat{X}_S(\Omega)$, reads out the atomic spin projection $\hat{X}_S$ at the rate $\Gamma_S\propto\Phi_1$. At the same time $\hat{H}_{\text{int},S}$ implies that measurement QBA due to $\hat{X}_{L\text{,in}}$ is imprinted on the atomic $\hat{P}_S$ quadrature. The atomic spin projection is driven in addition by intrinsic spin noise $\hat{{F}}_{S}$ so that ${\hat{X}}_{S}=\chi_{S}(\Omega) [\sqrt{\gamma_S}\hat{{F}}_{S} + \sqrt{\Gamma_S}\hat{{X}}_{L\mathrm{,in}}]$.  Here and henceforth we consider all quantities in Fourier (frequency) domain which is most appropriate for a  continuous-time measurement. The atomic oscillator's susceptibility $\chi_{S}(\Omega)=\pm 2\Omega_S/(\Omega_S^2-\Omega^2-2 i \Omega\gamma_S)$ is determined by the sign of its effective mass $(\pm)$, resonance frequency $\Omega_S$ and  relaxation rate $\gamma_{S}$ (half width at half maximum convention is used throughout the paper).  The physics of the QBA in the spin system can be understood as fluctuations of the Stark shift of the atomic energy levels due to fluctuations of the angular momentum of light  \cite{RMP}.

The classical drive LO$_1$ is filtered out after light passes through the atoms (Fig.~\ref{Fig:setup}a), whereas the relevant fluctuations in the orthogonal polarization, $\hat{P}_{L\text{,out}}^{S}$ and $\hat{X}_{L\text{,out}}^{S}=\hat{X}_{L\text{,in}}$ are mixed with a classical drive field LO$_2$ (with photon flux $\Phi_2$) in the same polarization and sent onto the mechanical oscillator. The phase of LO$_2$ is adjusted so that $\hat{X}_{L\text{,in}}^{M}=\hat{X}_{L\text{,out}}^{S}$, $\hat{P}_{L\text{,in}}^{M}=\hat{P}_{L\text{,out}}^{S}$. The linearized optomechanical Hamiltonian is $\hat{H}_{\text{int},M}\propto \hat{X}_{M}\hat{X}_{L\text{,in}}^{M}$ \cite{optomech}.  In analogy with the spin, the output phase quadrature of light, $\hat{P}_{L\text{,out}} =  \hat{P}_{L\text{,in}}^{M}+\sqrt{\Gamma_{M}}\hat{X}_{M}$, reads out the membrane position $\hat{X}_M$ at the rate $\Gamma_M\propto\Phi_2$. The membrane position is driven by thermal state noise $\hat{F}_{M}$ and the QBA of light, that is ${\hat{X}}_{M}=\chi_{M}(\Omega) [\sqrt{\gamma_{M0}}{\hat{F}}_{M}+\sqrt{\Gamma_{M}}{\hat{X}}_{L\mathrm{,in}}^{M}]$, where the mechanical susceptibility is given by $\chi_{M}(\Omega)=2\Omega_M/(\Omega_M^2-\Omega^2-2 i \Omega\gamma_{M0})$ and determined by the mechanical resonance frequency $\Omega_M$ and damping rate $\gamma_{M0}$. Hence $\hat{X}_{L\text{,in}}$ is the source of measurement QBA for both the membrane and the spin oscillator.

\begin{figure*}
\center\includegraphics[width=1.8\columnwidth]{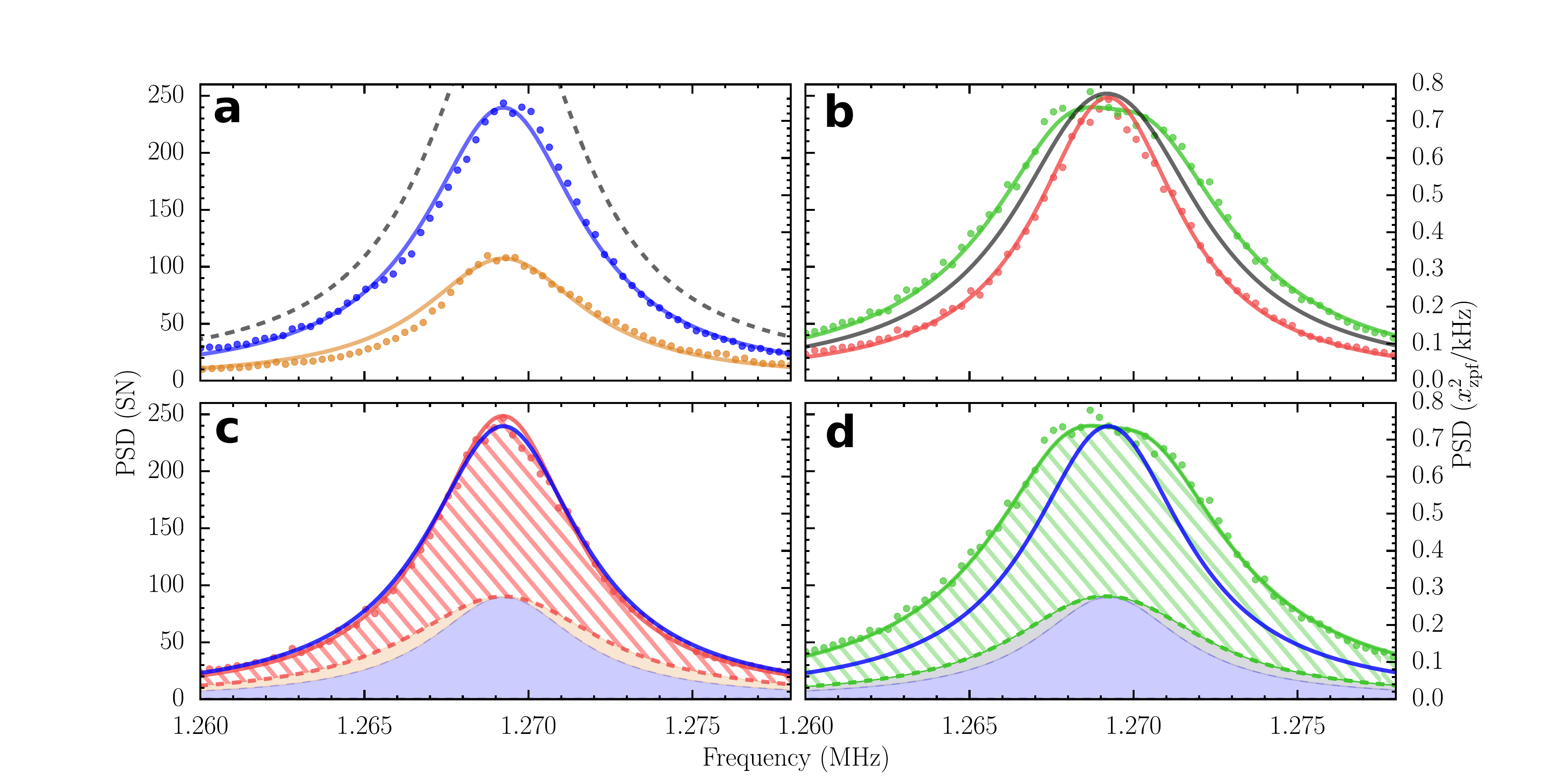}
\caption{
\textbf{Quantum back action for the mechanical and spin oscillators with equal central frequencies.} Axes labels: (SN) -- shot noise of light, $x_{\text{zpf}}$ -- zero point fluctuations.  A. Blue -- mechanical oscillator, brown -- spin oscillator, dashed -- the sum of the two spectra.  B. Hybrid
spectrum for the system with the negative (red) and positive (green) effective spin masses. Black curve
-- the model for the joint noise spectrum of the hybrid system with quantum BA interference put to zero. C. Hybrid spectrum noise for the negative mass (red dots). Thermal noise of the membrane (blue shade), thermal noise of the spin (brown shade) and joint thermal noise (red dashed curve). Striped area -- QBA of the hybrid system. Blue
curve -- model fit to the membrane noise data (same as in A). D. Same
as in C, but for the joint system with the positive mass spin oscillator. Curves -- full model (Supplementary Information).}\label{Fig:ZeroDet}
\end{figure*}

Overall, the homodyne readout of the joint system with the local oscillator LO$_3$ can be cast as
$\hat{P}_{L\text{,out}} =  \hat{P}_{L\text{,in}}+\sqrt{\Gamma_{M}}\hat{X}_{M}+\sqrt{\Gamma_{S}}\hat{X}_{S}$. The back action evading character of this measurement comes out most clearly when the measured light quadrature for the joint system is expressed as $\hat{P}_{L\text{,out}} =  \hat{P}_{L\text{,in}}+\sqrt{\Gamma_{M}\gamma_{M0}}\chi_{M}(\Omega)  \hat{F}_{M}+\sqrt{\Gamma_{S}\gamma_{S}}\chi_{S}(\Omega) \hat{{F}}_{S} + \left[\Gamma_{M}\chi_{M}(\Omega)+\Gamma_{S}\chi_{S}(\Omega)\right]\hat{{X}}_{L\mathrm{,in}}$, with the terms corresponding to shot noise of light, membrane thermal noise, spin noise, and  measurement QBA noise, respectively. Notably, the QBA term shows the interfering responses of the membrane and the spin oscillator. Ideal broadband QBA evasion is achieved for equal readout rates, $\Gamma_S=\Gamma_M$, and $\chi_{M}(\Omega)=-\chi_{S}(\Omega)$ which requires $\Omega_M=\Omega_S$, $\gamma_{M0}=\gamma_S$ and a negative mass spin oscillator (Supplementary Information and \cite{Bariani}).

We exploit the high level of flexibility in our modular hybrid setup to fulfill these requirements: It is straightforward to match the readout rates $\Gamma_M \simeq \Gamma_S$ by a proper choice of power levels $\Phi_{1,2}$, and to tune the atomic Larmor frequency $\Omega_S$ to the resonance frequency $\Omega_M=2\pi \times$\SI{1.28}{\MHz} of the mechanical drum mode. In order to observe appreciable QBA at the membrane's thermal environment of $7$K we use a phononic-bandgap shielded membrane with high mechanical quality factor $Q$ corresponding to an intrinsic damping rate of $\gamma_{M0}=2\pi\times$\SI{50}{\mHz}. On the other hand, the intrinsic spin damping rate $\gamma_{S0}\simeq 2\pi\times$\SI{500}{\Hz} is due to power broadening by optical pumping and atomic collisions. In addition, efficient spin readout requires significant power broadening by the probe light, $\gamma_S\gg\gamma_{S0}$ (Supplementary Information), impeding an adjustment of the spin to the mechanical linewidth. Instead we optically broaden the mechanical linewidth by introducing a detuning $\Delta<0$ of LO$_2$ from the cavity resonance.
This is a well established technique in optomechanical cooling experiments which exploits the dynamical back action of light on the mechanical oscillator for changing the mechanical susceptibility in order to generate a significantly enhanced effective damping rate $\gamma_M\gg\gamma_{M0}$ \cite{optomech}.
In this way matched linewidths $\gamma_M \simeq \gamma_S$ can be achieved by a proper choice of $\Phi_2$ and $\Delta$, cf.~Fig.~\ref{Fig:setup}c,d. The experimental parameters are listed in the Extended Data section. Introducing a nonzero detuning also modifies the optomechanical input-output relations and the QBA interference as detailed further below and in the Supplementary Information.

Having matched the susceptibilities and readout rates we perform a back-action limited readout of the two systems as shown in Fig.~\ref{Fig:setup}b,c,d. The ratio of QBA from vacuum noise of light $\hat{X}_{L,\mathrm{in}}$ to thermal noise due to $\hat{F}_{M(S)}$ is proportional to the quantum cooperativity parameters $C_{q}^{M(S)}$ respectively which we separately calibrate for each system (Supplementary Information). We achieve an optomechanical cooperativity of $C_{q}^M=2.5 \pm 0.3$ and on the side of atoms $C_{q}^S=1.10 \pm 0.15$ which signifies that QBA and thermal noise contribute roughly on the same level in both systems.

Fig.~\ref{Fig:ZeroDet} displays the results for the hybrid system. As a reference we show the spectra of the two individual systems taken separately (blue -- the mechanics, brown -- the spin) in Fig.~\ref{Fig:ZeroDet}a  both measured with the LO$_3$ detector. Fig.~\ref{Fig:ZeroDet}b presents the hybrid noise for the negative (red) and positive (green) effective spin masses, corresponding to two opposite orientations of the DC magnetic field relative to the spin polarization. The hybrid spectra  differ significantly from each other, with the area of the spectrum for the negative (positive) spin mass being significantly smaller (larger) than that for uncorrelated systems -- a clear demonstration of the destructive (constructive) interference of the QBA contributions for the two systems. We emphasize that these data signify a QBA cancellation irrespective of theoretical modelling. For comparison, the Fig.~\ref{Fig:ZeroDet}a also shows the curve (dashed) obtained by adding the two noise spectra recorded in separate measurements on atoms and the mechanical oscillator.

An intriguing feature of the hybrid noise spectra is the apparent absence of interference and noise cancellation exactly at the Fourier frequency $\Omega=\Omega_S=\Omega_M$ where the negative joint, positive joint and the mechanics spectra overlap  (Fig.~\ref{Fig:ZeroDet}b).  This is due to the strong optical broadening of the mechanical oscillator which leads to suppression of the spin phase noise contribution to light on the exact joint resonance. The effect is well understood from the full quantum model (Supplementary Information) and is analogous to optomechanically induced transparency ~\cite{Weis}. The solid red, green and blue curves for the negative joint, positive joint and mechanics, respectively, are generated from this model  and are in excellent agreement with the data. Fig.~\ref{Fig:ZeroDet}c presents the spectrum for the hybrid system with the negative mass and the model
fit (blue curve) to the spectrum of the mechanics (data in Fig.~\ref{Fig:ZeroDet}a). The noise reduction of
the hybrid spectrum (red dots) compared to the mechanics only (blue curve) in the wings of
the spectrum is observable, although its effect is diminished by the added spin thermal noise which
is present in the red data, but does not contribute to the blue curve. The observed variance
(area) for the joint negative system is $11.2\times x_{\text{zpf}}^2$ which is $(97\pm2)\%$ of the observed variance for the mechanical oscillator, where the error is derived from the fits. For the positive spin mass the constructive
interference of the QBA for the two systems is evident from comparing the green data points
to the blue curve for the membrane only (Fig.~\ref{Fig:ZeroDet}d).

\begin{figure*}
\center\includegraphics[width=1.8\columnwidth]{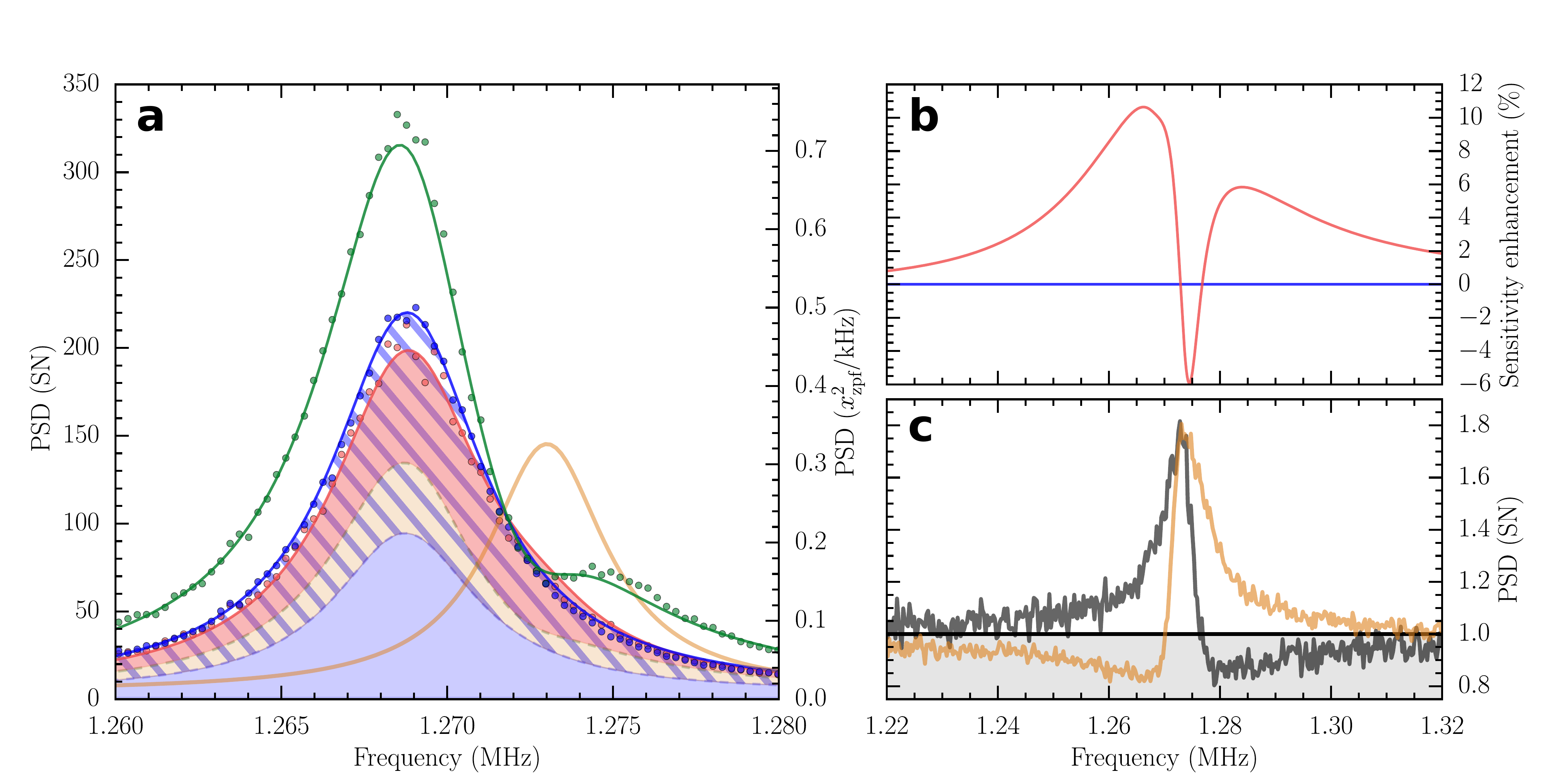}
\caption{
\textbf{Quantum back action for the optimally detuned mechanical and spin oscillators.} Noise spectra of detected light. Axes labels: (SN) -- shot noise of light, $x_{\text{zpf}}$ -- zero point fluctuations. 
A. Membrane noise (blue dots), hybrid system with the negative/positive mass spin oscillator (red/green dots), blue area -- membrane thermal noise, brown area -- spin thermal noise.  Solid brown curve -- a fit to the experimental spin spectrum taken without mechanical response. Red area -- QBA for the hybrid system. Striped area -- QBA for the membrane. B. Displacement sensitivity for the hybrid system with the negative mass (red) normalized to the sensitivity for the mechanical oscillator.  C. An example of the squeezed amplitude output of the spin system for positive (black) and negative (brown) effective mass.}\label{Fig:4kHzDet}
\end{figure*}

To find the reduction/enhancement of the QBA for the hybrid system, we use the calibration of the thermal noise described in the Supplementary Information and presented in Fig.~\ref{Fig:setup}c,d. The mechanical thermal noise found in Fig.~\ref{Fig:setup}d is shown as the blue shaded area in Fig.~\ref{Fig:ZeroDet}c,d. The spin thermal noise found in Fig.~\ref{Fig:setup}c is used as an input to the detailed model to find its contribution to the observed hybrid spectra (brown shaded area in Fig.~\ref{Fig:ZeroDet}c,d). Note that this noise is suppressed by the opto-mechanical response around $\Omega_M = \Omega_{S}$ by the same mechanism as the QBA contribution of the spin is reduced to zero at this point. Subtracting the thermal noise area from the total area, we find the QBA variance contribution for the hybrid negative system of $5.9 \times x_{\text{zpf}}^2$ (striped area in Fig.~\ref{Fig:ZeroDet}c) and for the hybrid positive system, $11.2 \times  x_{\text{zpf}}^2$ (striped area in Fig.~\ref{Fig:ZeroDet}d). Comparing these values with the QBA of $7.3 \times  x_{\text{zpf}}^2$ for the mechanical oscillator, we conclude that the variance of the QBA for the joint negative mass system is \SI{-1.0}{\dB} ($20\pm5\%$)  below the variance for the mechanics alone, whereas for the joint positive mass system it is \SI{1.7}{\dB} ($50\pm8\%$) higher. The main contributions to the error bars are the uncertainties in the calibration of quantum cooperativities.

Further studies reveal that a more efficient QBA evasion can be achieved when the two oscillator frequencies are not exactly equal,  $\Omega_M \neq \Omega_{S}$. Taking advantage of the straightforward tunability of  $\Omega_S$ with magnetic field, we run the QBA evasion experiment with the spin oscillator slightly detuned from the mechanical oscillator. In this case the best QBA evasion is obtained if the quadratures of light between the atomic and the optomechanical systems are rotated with respect to the phase of LO$_2$.  Fig.~\ref{Fig:4kHzDet}a shows the data for the hybrid system with the negative spin mass (red dots) with $\Omega_S - \Omega_M=2\pi \times 5.2$kHz and a phase rotation of $7$\textdegree, along with the noise of the mechanical oscillator (blue dots). For this experiment we find $C_q^M = 2.2$. We observe the broadband QBA evasion which, additionally, is most pronounced at $\Omega= \Omega_{M}$ where the mechanical response is maximal. The observed total variance for the hybrid system with the negative spin mass, $9.6\times x_{\text{zpf}}^2$, is $93\pm2\%$ of the variance for membrane only, $10.4\times x_{\text{zpf}}^2$. Note that interference in the hybrid system leads to suppression of the spin noise (solid brown curve) at $\Omega_{S}$, which is instead transformed into efficient QBA evasion around $\Omega_M$ for the negative mass hybrid system. Fig.~\ref{Fig:4kHzDet}b shows the improvement in the membrane displacement sensitivity obtained by the QBA evasion calculated as the ratio of the blue and red curves from Fig.~\ref{Fig:4kHzDet}a. These data signify broadband QBA evasion in a model independent way.

Subtracting thermal noise contributions we find the hybrid QBA (red area in Fig.~\ref{Fig:4kHzDet}a) of $4.1\times x_{\text{zpf}}^2$, that is \SI{-1.7}{\dB} ($32\pm5\%$) suppression compared to the mechanical QBA of $6.0 \times x_{\text{zpf}}^2$ (striped area). For the hybrid system with the positive spin mass (green dots), the QBA  is $10.3 \times x_{\text{zpf}}^2$ which is  \SI{2.4}{\dB} ($70\pm10\%$) above the QBA for the mechanics alone.
%a factor of $1.7\pm0.1$
In this detuned case the QBA reduction in case of negligible thermal noise can, in principle, overcome the limit of $1/2$ valid for the case of $\Omega_M=\Omega_S$ (see Supplementary Information), as indicated by the $60\%$ reduction of the classical BA that we have observed in an independent experiment with the system driven by classical white noise. The physics of the broadband QBA interference is due to the combination of the frequency dependent amplitude squeezing of the light generated by the spin and the interference of QBA of the two systems. An example of the amplitude squeezed output from the spin in shown in (Fig.~\ref{Fig:4kHzDet}c).

In conclusion, we have presented a novel hybrid quantum system consisting of distant mechanical and spin oscillators linked by propagating photons. Constructive or destructive interference of the quantum back action for the two oscillators depending on the sign of the effective mass of the spin oscillator is demonstrated. A detailed model describes the results with high accuracy. We have shown that the back action evading measurement in the hybrid system leads to the enhanced sensitivity of the displacement measurement. Further improvements are realistic with reduced propagation losses, even higher $Q$ mechanical oscillators \cite{membranes} and cavity enhanced spin systems. These results pave the way for entanglement generation and quantum communication between mechanical and spin systems, and to QBA free measurements of acceleration, gravity and force.

\bibliographystyle{naturemag}

\section*{Acknowledgments}
We acknowledge illuminating discussions with Farid Khalili. The cells with spin-protecting coating were fabricated by Mikhail Balabas. This work was supported by the European Union Seventh Framework Program (ERC grant INTERFACE, projects SIQS and iQUOEMS), the European Union's Horizon 2020 research and innovation programme (ERC grant Q-CEOM, grant agreement no. 638765), a Sapere Aude starting grant from the Danish Council for Independent Research, and the DARPA project QUASAR. R.A.T. is funded by the program Science without Borders of the Brazilian Federal Government. E.Z. is supported by the Carlsberg Foundation. We acknowledge help from Marius Gaudesius at the early stage of the experimental development.

\section*{Author Contributions}
E.S.P. conceived and led the project. C.B.M, R.A.T and G.V. built the experiment with the help of K.J., Y.T. and A.S. The membrane resonator has been designed and fabricated by Y.T. C.B.M, R.A.T, G.V. and E.S.P. took the data. E.Z. and K.H. developed the theory with input from A.S. and E.S.P. The paper was written by E.S.P., K.H., E.Z., R.A.T., C.B.M. and G.V. with contributions from other authors. A.S., K.H. and E.S.P supervised the research.

\end{document}

\title{Supplementary Information for \\
``Quantum back action evading measurement of motion\\in a negative mass reference frame''}
\author{Christoffer B. M{\o}ller}
\author{Rodrigo A. Thomas}
\affiliation{Niels Bohr Institute, University of Copenhagen, DK-2100 Copenhagen,
Denmark}
\author{Georgios Vasilakis}
\affiliation{Niels Bohr Institute, University of Copenhagen, DK-2100 Copenhagen,
Denmark}
\affiliation{Institute for Electronic Structure and Laser, Foundation for Research and Technology-Hellas, Heraklion 71110, Greece}
\author{Emil Zeuthen}
\affiliation{Institute for Theoretical Physics \& Institute for Gravitational
Physics (Albert Einstein Institute), Leibniz Universit\"{a}t Hannover,
Callinstra{\ss}e 38, 30167 Hannover, Germany}
\affiliation{Niels Bohr Institute, University of Copenhagen, DK-2100 Copenhagen,
Denmark}
\author{Yeghishe Tsaturyan}
\author{Kasper Jensen}
\author{Albert Schliesser}
\affiliation{Niels Bohr Institute, University of Copenhagen, DK-2100 Copenhagen,
Denmark}
\author{Klemens Hammerer}
\affiliation{Institute for Theoretical Physics \& Institute for Gravitational
Physics (Albert Einstein Institute), Leibniz Universit\"{a}t Hannover,
Callinstra{\ss}e 38, 30167 Hannover, Germany}
\author{Eugene S. Polzik}
\affiliation{Niels Bohr Institute, University of Copenhagen, DK-2100 Copenhagen,
 Denmark}

\maketitle

\section{Experimental setup}

\subsection{Atomic spin oscillator}

The spin ensemble consists of $N_a \sim 10^{9}$ Cesium atoms contained in an anti-relaxation coated pyrex vapor cell (microcell) \cite{MishaC30} heated to a temperature of \SI{65}{\degreeCelsius}. These atoms are confined in a channel of $300\si{\micro\meter} \times 300\si{\micro\meter} \times 10\si{\milli\meter}$ connected to an external Cesium reservoir via a $\sim 10\si{\micro\meter}$ radius laser drilled hole, as shown in Figure 1b. They are addressed by light of a waist (radius) size of \SI{55}{\micro\meter} focused through the channel. The microcell is enclosed in a four layer magnetic shielding, protecting the spins from ambient magnetic fields and external RF sources. An inner system of coils produces a homogeneous bias field, $B$, which leads to a Larmor frequency $\Omega_S$. The wall-to-wall transient time of Cs atoms in the channel is on average $\simeq$ \SI{1.5}{\micro\second}. Within the characteristic evolution time of the quantum state, the moving atoms cross the light beam many times and thus experience a motionally averaged interaction \cite{RMP,Borregaard}. The spin-protecting coating of the cell walls grants an intrinsic decoherence rate of $\simeq$\SI{300}{\micro\second}. This rate is limited by spin destruction collisions of atoms with the cell walls, magnetic field inhomogeneities and spin-exchange collisions.

A circularly polarized  diode laser tuned to the $F=3 \rightarrow F'=4$ transition of the D2 line is used to spin polarize atoms into the $F=4$ ground state. The probing of the atomic ensemble is done with linearly polarized light at \SI{852.3490}{\nano\meter} ($\text{LO}_1$), blue detuned by $\Delta_S \sim 2\pi \times$\SI{3}{\giga\hertz} from the $F=4 \rightarrow F'=5$ in the $\text{D}_2$ transition. Absorption effects can be neglected as $\Delta_S \gg \Delta \omega_\text{Doppler}^{\text{HWHM}}\simeq 2\pi \times$ \SI{200}{\mega\hertz}, effectively eliminating the effect of spin motion. The polarization of the probe set by the half-waveplate (HWP$_0$ in Fig.~\ref{fig:DetailedExperimentalSetup}) allows for adjustment of the polarization axis of the linearly polarized probing light (LO$_1$). In the experiment this axis is chosen to minimize the added broadening of the spin oscillator, resulting in an angle $\theta_a \approx 55^{\circ}$ with respect to the direction of the atomic polarization.  We emphasize that the dominant part of the light-atoms interaction is of the QND type~\cite{Comment2} (see Eq.~\eqref{eq:HamiltonianSpin0}) and does not depend on the angle $\theta_a$; the optical rotation that the light experiences is to a very good approximation independent of the polarization orientation. The vacuum sidebands that affect the spin oscillator are in an orthogonal polarization mode and $\pi/2$ out of phase with respect to the local oscillator LO$_1$.

The optical readout rate, $\Gamma_S \propto \Delta_S^{-2} \Phi_1 |J_x|$, is a function of the number of atoms $J_x = F N_a$, the local oscillator photon flux $\Phi_1$ and the detuning $\Delta_S$. Its origin lies in the re-parametrization of the Faraday rotation experienced by light due to interaction with far detuned atoms \cite{Kasperthesis,Vasilakis2015}. In the established language of the light-atoms interface \cite{RMP}, the readout rate is related to the interaction strength $\kappa^2_{\text{atoms}} = \Gamma_S T$, with $T$ being the temporal length of the probing light mode. The atomic spins' linewidth, $\gamma_S$, is dominated by the optical broadening and is proportional to $\Phi_1$ in the regime of interest. Typical values for optical powers and $\gamma_S$ are presented in Table~\ref{table:parameters}.

The Cesium spin ensemble fully polarized to $F=4, m_F=\pm 4$ has projection noise variance $\text{var}(\hat{J}_{y,z})_{\text{PN}}=|J_x|\hbar/2=FN_A\hbar/2 = 2N_A \hbar$ in its ground state, whereas a completely unpolarized ensemble has $\text{var}(\hat{J}_{y,z})_{\text{Th}}=F(F+1)N_A\hbar/3=10/3\times \text{var}(\hat{J}_{y,z})_{\text{PN}}$ \cite{RMP}. In the experiment $\text{var}(\hat{J}_{y,z})\simeq 2.8\times \text{var}(\hat{J}_{y,z})_{\text{PN}}$ with the degree of spin polarization of ~$60\%$, equivalent to having $1.8$ units of extra ground state noise. The negative (positive) mass configuration is achieved by optical pumping of the atoms to the $F=4, m_F=+4$ ($F=4, m_F=-4$) state, i.e, parallel (antiparallel) to the magnetic field, which we take to define the positive $x$-direction. Within the Holstein-Primakoff approximation \cite{HP} this is formally equivalent to having a harmonic oscillator with  $\Omega_S < 0$ ($\Omega_S > 0$) as depicted in Figure 1c (Figure 1d) of the main text. Note that throughout the Methods, we include the sign of the effective spin mass in $\Omega_S$ (whereas this sign is stated as an explicit prefactor in the main text). For the negative (positive) mass oscillator, the first excited state of the spin oscillator is the one with a single atom in the $F=4, m_F=3$ ($F=4, m_F=-3$) state. Experimentally, we change the magnetic field direction to choose the sign of the oscillator's mass.

As presented in Figure 2c, the quantum cooperativity for the spin oscillator, $C_q^S$, is characterized via broadband thermal modulation of the optical driving force. This technique requires an electro-optical modulator to drive LO$_1$'s polarization quadrature $\hat{X}_{L,\text{in}}$ and a RF source outputing thermal voltage noise, producing a frequency independent and proportional to the driving voltage, optical power with excess $n_\text{WN}$ photon flux in the frequency band of interest. In Figure 2c, the black dots represent the spin oscillator driven with $n_\text{WN}=1.22$. Comparing this curve to the shot-noise limited ($n_\text{WN}=0$) probing, in brown dots, we extract the effect of quantum noise of light on the spin oscillator and the ratio of back action to thermal noise, the quantum cooperativity. For a thorough discussion on this we refer to the Section~\ref{sec:calibrationspins}. 

\subsection{Optomechanical System}

The optomechanical system is based on a near-monolithic cryogenic membrane-in-the-middle system described in detail elsewhere \cite{Nielsen}.
The mechanical oscillator is a highly stressed, $60~\si{nm}$ thick SiN membrane supported by a silicon periodic structure forming a phononic bandgap. The bandgap protects the oscillator from phonon tunneling from the clamping of the structure and provides a region clear of undesired phononic modes \cite{Yeghishe}. The high stress boosts the quality factor through dissipation dilution [28]. The membrane thickness, $t_M$, is chosen to maximize the optomechanical single photon coupling rate, $g_0$. This introduces a tradeoff between the zero point fluctuations $x_{\text{zpf}}=\sqrt{\hbar/(m_{\text{eff}} \Omega_M)} \propto t_M^{-1/2}$ and the amplitude reflection coefficient of SiN which is periodic with $t_M$. 

The (1,2) drum mode of the membrane with frequency $\Omega_M=2\pi \times 1.28~\si{MHz}$ is used as it is the lowest frequency mode to lie within the bandgap and has a high quality factor of $Q=13 \times10^6$ as measured by ring-downs ($\gamma_{M0} = 2\pi \times 50~\si{mHz}$). The $\sim 8\%$ side length difference of the membrane breaks the degeneracy of the (1,2) and (2,1) modes significantly, with the (1,2) mode being $\sim 60\si{kHz}$ lower in frequency than its sibling. This membrane is placed in a cavity and aligned such that the cavity TEM$_{00}$ mode has a good overlap with the (1,2) mode and a poor overlap with the (2,1) mode. This further separates the systems as the optical spring effect (dynamical back action) pushes the (1,2) mode another $\sim 10\si{kHz}$ away, while having only a marginal impact on the (2,1) mode. 

The $1.3~\si{mm}$ long plano-concave Fabry-P\'erot optical cavity with finesse $F=4500$ (half bandwidth of $\kappa=2\pi \times 13~\si{MHz}$) is mounted in a continuous flow cryostat with large windows for good optical access and a base temperature of $4.4~\si{K}$. The power tranmissions of the mirrors are $20~\si{ppm}$ and $1400~\si{ppm}$ thus giving a largely one-sided cavity. Placing the aforementioned dielectric membrane $500~\si{\micro m}$ from the $20~\si{ppm}$ mirror, forms two sub-cavities whose dynamics can be mapped onto the canonical optomechanical formulation used in the theory section via the transfer matrix model approach described in \cite{Jayich2008}. In effect, the cavity half bandwidth $\kappa$ is modulated depending on the position of the membrane with respect to the standing wave in the cavity \cite{Wilson2009}. This is due to a differential intracavity photon population being built up in each sub-cavity. Having more intracavity photons populating the sub-cavity bounded on one side by the low transmission ($20~\si{ppm}$) mirror produces an overall decreased cavity loss rate. In the canonical formulation this is equivalent to the decay rates $\kappa_1$ and $\kappa_2$ of the cavity ports 1 and 2 being altered asymmetrically, see  Fig.~\ref{fig:TheorySchematic}. The membrane itself adds negligible additional loss.

We position the membrane in the cavity such that the coupling rate is large and where the overall cavity bandwidth is reduced from $\kappa = 2\pi \times 13\si{MHz}$ (bare cavity) to $\kappa = 2\pi \times 7.7\si{MHz}$ (finesse enhanced to $7500$). This comes at the expense of having a less one-sided cavity. The ratio of the cavity ports decay rates goes from $\kappa_1 / \kappa_2 = 70$ (bare cavity) to $\kappa_1 / \kappa_2 = 25$. The reduced cavity bandwidth is advantageous as a certain degree of sideband resolution is required to optically broaden the mechanical oscillator significantly without requiring a too large readout rate (required to match the spin system). 

The high transmission incoupling mirror is mounted on a piezoelectric transducer, which is used to tune the cavity resonance frequency close to that dictated by the atomic probe LO$_1$. It is tuned such that LO$_2$ probes the cavity red detuned by $\Delta=-2\pi \times 4.7~\si{MHz}$. This is ensured using a separate beam originating from the same laser. This beam is blue shifted (by $-\Delta$) from LO$_2$ described above by an acousto-optic modulator. It is then phase modulated at 12MHz and probes the cavity from the undercoupled port 2, see Fig.~\ref{fig:DetailedExperimentalSetup}. An error signal is derived using the Pound-Drever-Hall technique and is used to feedback on the aforementioned piezoelectric transducer which stabilizes the cavity such that this beam is locked on resonance. This locking beam is in the orthogonal polarization to LO$_2$ and contributes to $<1\%$ of the intracavity power. The locking beam thus has a negligible impact on the intracavity dynamics and the final detection. 

The mechanical oscillator is initially only coupled to a thermal bath of temperature $T_{\text{bath}}$ with mean occupation $\bar{n}_{\text{bath}}$. Adding the probe LO$_2$ alters the dynamics of the system as the oscillator is coupled to the intracavity field with a rate $g = g_0 \sqrt{N}$, where $N$ is the mean intracavity photon number. Dynamical back action optically broadens the mechanical linewidth by $\gamma_{M,opt}$ (such that $\gamma_M = \gamma_{M0} + \gamma_{M,opt}$) and the mean thermal occupation is reduced to $\bar{n}_M^{th} = (\gamma_{M0}/\gamma_M) \bar{n}_{\text{bath}}$. This (so-called) sideband cooling is due to an asymmetry in the Stokes and anti-Stokes processes caused by the detuning from cavity resonance. The Stokes sideband (causing heating) is never completely suppressed which sets a minimum achievable mean occupation $\bar{n}_{M}^{min}$ \cite{Marquardt2007}. The effective mean occupation of the mechanical oscillator is now $\bar{n}_{M} = (\gamma_{M0}/\gamma_M)\bar{n}_{\text{bath}} +(\gamma_{M,opt}/\gamma_M) \bar{n}_M^{min}$, where the contribution $\bar{n}_M^{min}$ is correlated with the quantum back-action. 

The total variance of motion will thus have contributions from both the QBA and the thermal bath; the ratio of these is
\begin{equation}
\frac{\gamma_{M,opt}(\bar{n}_M^{min} + 1/2)}{\gamma_{M0}(\bar{n}_{\text{bath}} + 1/2)} \approx \frac{C_q^M}{2} \left( \frac{\kappa^2}{\kappa^2 + (\Delta - \Omega_M)^2} + \frac{\kappa^2}{\kappa^2 + (\Delta + \Omega_M)^2}\right), \label{eq:BA2TN}
\end{equation}
where we have introduced the quantum cooperativity $C_q^M = g_0^2 N/(2\kappa \gamma_{M0}\bar{n}_{\text{bath}})$ and approximated $\bar{n}_{\text{bath}} + 1/2 \approx \bar{n}_{\text{bath}}$ (in the present scenario $\bar{n}_{\text{bath}}\sim10^5$).

The total variance of the motion can be directly inferred from the area of the measured output spectrum by $\langle\hat{X}_M^2\rangle = \int (\Gamma_M^{-1}\langle\hat{P}_{L\text{,out}}^2\rangle - 1)d\Omega \approx \Gamma_M^{-1}\int \langle\hat{P}_{L\text{,out}}^2\rangle d\Omega$. The approximation is good as the spectra used are dominated by QBA and thermal noise contributions, with a negligible SN contribution, in the frequency range shown in all Figures. By the same token, Eq.~\eqref{eq:BA2TN} gives operational meaning to the quantum cooperativity $C_q^M$ as the ratio of QBA and thermal contributions to the observed variance (except for the Lorentzian factor of order unity, $\lesssim 1$, seen on the right-hand side).

The optomechanical system is probed in reflection. The combination of a quarter-waveplate at $45^{\circ}$ and a polarizing beam splitter effectively acts as an isolator transmitting the input light (in the hybrid configuration coming from the spin system) and reflecting the light emerging from the cavity. With well characterized optical losses and system parameters, the bath temperature and $C_{q}^M$ can be inferred from the observed degree of ponderomotive squeezing. A spectrum (amplitude quadrature) showing squeezing of -\SI{1.7}{\dB} (-\SI{2.6}{\dB} corrected for detection efficiency of $72\%$) is shown in Figure 2b. The shot noise (SN) level is verified by balanced detection and by comparison to a white light source to within $<5\%$ accuracy. Using the detailed model (outlined in the theory section) with the bath temperature of the membrane, $T_\text{bath}$, as the only adjustable parameter, we obtain the fit shown in Figure 2b with $T_\text{bath}=(7\pm.5)~\si{K}$. From $T_\text{bath}$ we thus obtain the value of $C_{q}^M=2.5\pm0.3$ for the data presented in Figure 2d. The thermal noise contribution is shown as the blue area and can easily be found using Eq.~\eqref{eq:BA2TN}.
  
For the two phase quadrature data sets shown the mechanical oscillator was optically broadened to $\gamma_M=2\pi \times2.7~\si{kHz}$. This required the input LO$_2$ power to be adjusted as the cavity half bandwidth and single photon coupling rate varied slightly between runs. This was due to a varying membrane position with respect to the intra-cavity field on different days caused by overnight temperature cycles. The experimentally realised parameters for these are displayed in Table~\ref{table:parameters}. 

\subsection{Hybrid system}

\begin{figure}
\includegraphics[width=1\columnwidth]{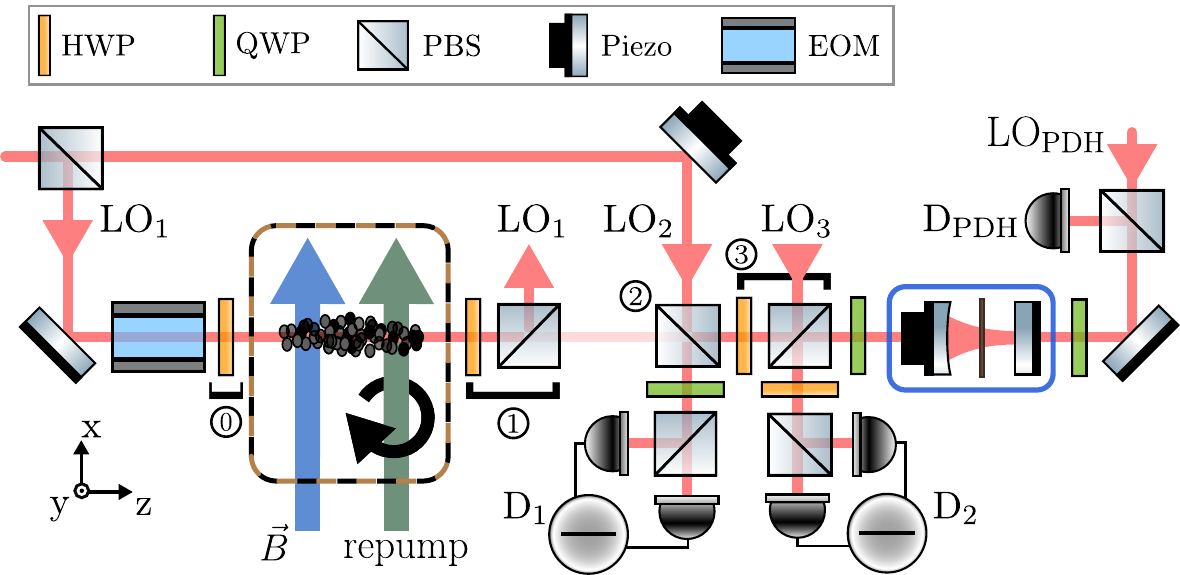}
\caption{Detailed schematic of the experimental setup. The atomic spin system is pictured in the black-brown dashed box, along with its B-field and optical pumping; in the blue lined box, the optomechanical membrane-in-the-middle setup. The hybrid system is probed via a travelling optical mode. The atomic system, driven by LO$_1$ with linear polarization angle set by \raisebox{.5pt}{\textcircled{\raisebox{-.9pt} {0}}} has its output polarization filtered in \raisebox{.5pt}{\textcircled{\raisebox{-.9pt} {1}}} and is recombined with the correct phase with LO$_2$ in \raisebox{.5pt}{\textcircled{\raisebox{-.9pt} {2}}}, set electronically via suitable detection in D$_1$; \raisebox{.5pt}{\textcircled{\raisebox{-.9pt} {3}}} ensures that both local oscillator and the filtered atomic response have the same polarization. The optomechanical system is probed in reflection and frequency stabilized via PDH locking in the unused port. Phase sensitive detection is done via homodyning with LO$_3$ in D$_2$.}
\label{fig:DetailedExperimentalSetup}
\end{figure}

A detailed schematic of the experimental setup is shown in Fig.~\ref{fig:DetailedExperimentalSetup}: a travelling light field interacts with the atomic spin and the mechanical oscillator in a cascaded way. The driving beams for the system, LO$_1$ and LO$_2$, as well as the local oscillator LO$_3$ for the homodyne measurement are generated by a Ti-Sapphire laser. The light is shot noise limited for the relevant powers and Fourier frequencies of interest in the quadratures that matter for both systems. 

The spin-mechanics interface requires filtering of the spins' output field. The atomic spins respond to modulation out of phase and in the orthogonal polarization to its local oscillator (denoted by the quantum field operator $\hat{X}_{L,\text{in}}$ in the main text). The mechanical oscillator responds to modulation in phase and in the same polarization mode as its own driving local oscillator (represented by the field operator $\hat{X}^{M}_{L,\text{in}}$). Therefore, LO$_2$ should be in an orthogonal polarization mode and $\pi/2$ phase shifted with respect to LO1. Two filtering stages are required, one for polarization and one for phase, both of them being depicted in Fig.~\ref{fig:DetailedExperimentalSetup} and described in the following. The polarization filtering is done using HWP$_1$ and PBS$_1$ right after the microcell, decoupling LO$_1$ from the quantum fluctuations of interest in the orthogonal polarization quadrature; the phase filtering is realized using a Mach-Zehnder interferometer with output at PBS$_2$, setting a variable phase for the spin sideband fields  $\{\hat{X}^{S}_{L,\text{out}},\hat{P}^{S}_{L,\text{out}}\}$ in respect to LO$_2$. The driving local oscillator and the sidebands are then projected in the same polarization mode with HWP$_3$ and PBS$_3$ and coupled to the optomechanical cavity. To detect the optical quadrature of interest, balanced polarimetry with the local oscillator LO$_3$ (with the aid of another Mach-Zehnder interferometer) is performed when the phase quadrature is of interest; to measure of the amplitude quadrature, the local oscillator is removed and all light is directed to a single photodiode.

In the experiment, PBS$_1$ extinguishes LO$_1$ better than $1$:$10^3$ from the optical path with little loss of the modulation sidebands or those carrying information about the spin oscillator. 

An electro-optic modulator (EOM) in LO$_1$ is used for locking the phase filtering interferometer.  The phase and axis of the EOM are adjusted in such a way so that a voltage modulation (small with respect to the $\pi$ voltage) results in a small modulation predominantly in the degree of circular polarization of light (quantified by the $\hat{S}_z$ Stokes component). The ratio of circular polarization modulation to linear polarization modulation introduced by the EOM is typically $\sim10^5$ in power. Sinusoidal modulation sidebands at $10\si{MHz}$, far from both oscillators' responses, provide the phase reference. These sidebands are combined with LO$_2$ in the output of the interferometer, PBS$_2$. A half-waveplate allows for an adjustable fraction of the sideband power to be used for locking (typically $\sim 5\%$). The demodulated result of the balanced polarimetry measurement of the locking signal in D$_1$ is proportional to $\cos{\delta \phi_{\text{LO1,2}}}$, where $\delta \phi_{\text{LO}_{1,2}}$ is the phase difference between LO$_1$ and LO$_2$. Feedback on a piezoelectric transducer proportional to this signal allows us to lock the new local oscillator in phase with the sideband quadrature that drove the spin oscillator. 

The half-waveplate HWP$_2$ and PBS$_2$ project a small portion of LO$_2$ ($\sim 3\%$) and most of the sideband signal ($\sim 97\%$) into the same polarization mode. Suitable optics direct the beam onto the optomechanical cavity with a total optical power transmission for the spin system response sidebands in the $60-65\%$ range. These sidebands are in the same spatial mode as LO$_1$, which is modematched to the cavity with an efficiency of $\eta_{mm}\simeq 90\%$. The modematching is defined as the fraction of incident LO$_1$ power going into the TEM$_{00}$ compared to all TEM modes. 

When the characterization of the atomic spin oscillator is performed, a function generator provides a white noise (WN) modulation over the interesting frequency range, from \si{1\mega\hertz} to \si{2\mega\hertz}; typical values for the added modulated WN photons (in units of SN) ranges from 0.5 to 100. 

The spectra of $\hat{P}_{L,\text{out}}$ are measured by balanced homodyning of the field reflected off the optomechanical cavity with LO$_3$, with power in the order of \SI{1.5}{\mW}, which is locked to the DC zero of the interference fringe with LO$_2$, thus ensuring that the phase quadrature is being measured.

The model fits and knowledge of all relevant system parameters provide a reliable reference point from which we calibrate the spectra in units of the mechanical zero point fluctuations. For example, the right vertical axis in Fig.~1d shows the spectral density of motion in units of $x_{\text{zpf}}^2/\si{kHz}$, calibrated by dividing the thermal noise by $n_{\text{th}}$. Integrating the power spectral density data we find the observed membrane variance of $11.6 \times x_{\text{zpf}}^2$. Subtracting the thermal noise variance $4.3 \times x_{\text{zpf}}^2$ we obtain the QBA variance $7.3 \times x_{\text{zpf}}^2$.

Losses in the system are due to the finite transmission coefficient between the spin and the optomechanical systems of $\eta_1=0.53$, which includes the finite optomechanical coupling efficiency $0.88$, and the transmission coefficient between the optomechanical system and the detection of $\eta_2=0.64$ which includes the quantum efficiency of the photodetector of $0.90$. These values vary within a few percentage points from experiment to experiment.

\section*{Theoretical model}

\subsection{Optomechanical System}
\label{OptomechanicalTheorySection}

\begin{figure}[h]
  \centering
  \includegraphics[width=\textwidth]{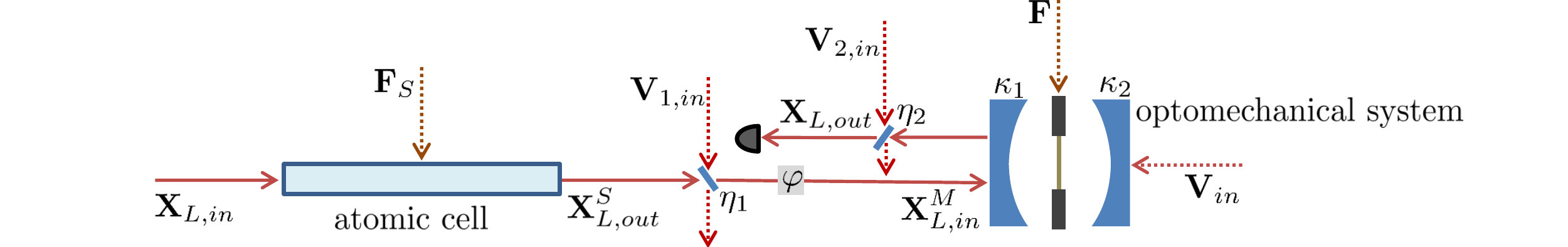}
  \caption{Schematic of the setup: The atomic spin is driven by light noise ${\bf X}_{L,in}$ and spin noise ${\bf F}_S$. Output light of the the spin system ${\bf X}^S_{L,out}$ is channeled to the atomic system, and experiences losses characterized by a transmissivity $\eta_1$ associated with additional light noise ${\bf V}_{1,in}$ and a phase rotation by an angle $\varphi$, resulting in a driving field ${\bf X}^M_{L,in}$ of the optomechanical system. The optomechanical cavity is two-sided with decay rates $\kappa_1$ and $\kappa_2$. The optomechanical system is driven in addition by light noise ${\bf V}_{in}$ and a thermal force ${\bf F}$. The output field of the optomechanical system experiences further losses with transmissivity $\eta_2$ associated with additional light noise ${\bf V}_{2,in}$.}
  \label{fig:TheorySchematic}
\end{figure}

%\begin{itemize}
   The optomechanical system is described by the standard linearized Hamiltonian
  \begin{align*}
  \hat{H}
  &=\frac{\Omega_M}{2}\left(\hat X_M^2+\hat P_M^2\right) - \Delta\, \hat a^\dagger \hat a - g \left(\hat a e^{i\phi}+\hat a^\dagger e^{-i\phi}\right)\hat X_M,
  \end{align*}
  where $[\hat X_M,\hat P_M]=i$ are the dimensionless canonical operators for the mechanical system, and $[\hat a,\hat a^\dagger]=1$ are annihilation/creation operators for cavity photons. $\Delta=\omega_L-\omega_c$ is the detuning of the driving laser from the cavity resonance $\omega_c$. The linearized optomechanical coupling rate $g=g_0|\alpha|$ depends on the single photon coupling rate $g_0$ of the optomechanical system and the intracavity amplitude $\alpha$. It is linked to the optomechanical readout rate $\Gamma_M$ introduced in the main text by
   \begin{equation}
 \Gamma_M=\frac{2g^2}{\kappa},\label{eq:Gamma_M}
\end{equation}
   where $\kappa$ is the cavity half linewidth. The phase $\phi=\arctan(\Delta/\kappa)$ denotes the phase of the intracavity field amplitude $\alpha$ relative to driving field, as is discussed further below in Eq.~\eqref{eq:phi}.  Here we take the incoming field as the phase reference instead  of the intracavity field (as is usually done in cavity optomechanics)  since we eventually interested in the transfer matrix for the incoming/outgoing amplitudes resulting from this Hamiltonian.
   Including decay and Langevin noise forces the equations of motion corresponding to the Hamiltonian are

  \begin{align*}
    \dot{\hat a}(t)+(\kappa-i\Delta)\hat  a(t)-ig e^{-i\phi} \hat X_M(t)&=\sqrt{2\kappa_1}\hat a_{in}(t)+\sqrt{2\kappa_2}\hat v_{in}(t)\\
    \ddot{\hat X}_M(t)+\Omega^2_M \hat X_M(t)+2\gamma_{M0} \dot{\hat X}_M(t)-\Omega_M g\left(\hat a(t) e^{i\phi}+\hat a^\dagger(t) e^{-i\phi}\right)&=\sqrt{4\gamma_{M0}}\,\Omega_M \hat f(t),
  \end{align*}
where $\hat a_{in}(t)$ and $\hat v_{in}(t)$ are incoming quantum fields driving the cavity through port $1$ and $2$, respectively, cf.~Fig.~\ref{fig:TheorySchematic}; their commutation relations are $[\hat{a}_{in}(t),\hat{a}^{\dagger}_{in}(t')]=\delta(t-t')=[\hat{v}_{in}(t),\hat{v}^{\dagger}_{in}(t')]$.
The partial decay rates $\kappa_{1(2)}$ fulfill $ \kappa=\kappa_1+\kappa_2$. The linewidth of the mechanical resonance (excluding optical broadening) is $\gamma_{M0}$, and the thermal Langenvin force is $\hat f(t)$. In the high temperature limit we can take $\langle \hat{f}(t)\hat{f}(t')\rangle=\delta(t-t')(\bar{n}+1/2)$ where $\bar{n} \simeq k_{B}T_{\text{bath}}/(\hbar\Omega_M)$.

   In the frequency domain the equations of motion read
  \begin{align*}
    \big(\kappa-i(\Delta+\Omega)\big){\hat a}(\Omega)-ig e^{-i\phi} \hat X_M(\Omega)&=\sqrt{2\kappa_1}\hat a_{in}(\Omega)+\sqrt{2\kappa_2}\hat v_{in}(\Omega)\\
   D_{M0}(\Omega) {\hat X}_M(\Omega)-g\,\Omega_M\left(\hat a(\Omega) e^{i\phi}+\hat a^\dagger(-\Omega) e^{-i\phi}\right)&=\sqrt{4\gamma_{M0}} \Omega_m\hat f(\Omega),
  \end{align*}
where
\begin{align}
  D_{M0}(\Omega)&=\Omega_{M}^2-\Omega^2-2i\Omega\gamma_{M0}.\label{eq:D_M0}
\end{align}
We define field quadratures as
\begin{align}
  \hat X_L(\Omega)&=\frac{1}{2}(\hat a(\Omega)+\hat a^\dagger(-\Omega)) & \hat P_L(\Omega)&=\frac{1}{2i}(\hat a(\Omega)-\hat a^\dagger(-\Omega))\label{eq:2phot-quad}
\end{align}
and similar definitions are used for quadratures of incoming/outgoing fields.  In terms of these the equations of motion in frequency domain are
\begin{align*}
  \begin{pmatrix}
  \kappa-i\Omega & \Delta & -g\sin\phi \\
  -\Delta & \kappa-i\Omega & -g\cos\phi \\
  -2g\Omega_M\cos\phi & 2g\Omega_M\sin\phi & D_{M0}(\Omega)
  \end{pmatrix}
  \begin{pmatrix}
    \hat X_L(\Omega) \\ \hat P_L(\Omega) \\ \hat X_M(\Omega)
  \end{pmatrix}&=
  \begin{pmatrix}
    \sqrt{2\kappa_1} \hat X^M_{L,in}(\Omega)+\sqrt{2\kappa_2}\hat V_{x,in}(\Omega) \\
    \sqrt{2\kappa_1}\hat P^M_{L,in}(\Omega)+\sqrt{2\kappa_2}\hat V_{p,in}(\Omega) \\
    \sqrt{4\gamma_{M0}}\Omega_M \hat f(\Omega)
  \end{pmatrix}
\end{align*}
which can be conveniently written in terms of block matrices
\begin{align*}
  \begin{pmatrix}
    O^T_\phi & 0 \\ 0 & 1
  \end{pmatrix}
  \begin{pmatrix}
    A & B \\ C & D_{M0}(\Omega)
  \end{pmatrix}
  \begin{pmatrix}
    O_\phi & 0 \\ 0 & 1
  \end{pmatrix}
  \begin{pmatrix}
    {\bf \hat X}_L(\Omega) \\ \hat X_M(\Omega)
  \end{pmatrix}&=
  \begin{pmatrix}
    \sqrt{2\kappa_1}{\bf \hat X}^M_{L,in}(\Omega)+\sqrt{2\kappa_2}{\bf \hat V}_{in}(\Omega)\\
    \sqrt{4\gamma_{M0}}\Omega_M \hat f(\Omega)
  \end{pmatrix}
\end{align*}
where
\begin{gather*}
O_\phi=\begin{pmatrix} \cos(\phi) & -\sin(\phi) \\ \sin(\phi) & \cos(\phi) \end{pmatrix},\quad 
A=\begin{pmatrix} \kappa-i\Omega & \Delta \\ -\Delta & \kappa-i\Omega \end{pmatrix},\\
B=\begin{pmatrix} 0 \\ - g \end{pmatrix},\quad 
C=\begin{pmatrix} -2g\Omega_M & 0 \end{pmatrix},\quad 
{\bf \hat X}_L(\Omega)=\begin{pmatrix}\hat  X_L(\Omega) \\ \hat P_L(\Omega) \end{pmatrix},\quad\textrm{etc.}
\end{gather*}
 The equations of motion are solved by
\begin{align*}
  \begin{pmatrix}
    {\bf \hat X}_L(\Omega) \\ \hat X_M(\Omega)
  \end{pmatrix}&=
  \begin{pmatrix}
    O^T_\phi & 0 \\ 0 & 1
  \end{pmatrix}
  \begin{pmatrix}
    A & B \\ C & D_{M0}(\Omega)
  \end{pmatrix}^{-1}
  \begin{pmatrix}
    O_\phi & 0 \\ 0 & 1
  \end{pmatrix}
  \begin{pmatrix}
    \sqrt{2\kappa_1}{\bf \hat X}^M_{L,in}(\Omega)+\sqrt{2\kappa_2}{\bf \hat V}_{in}(\Omega)\\
    \sqrt{4\gamma_{M0}}\Omega_M \hat f(\Omega)
  \end{pmatrix}
\end{align*}
where the inverse Block matrix can be expressed in two equivalent forms
\begin{align}
  \begin{pmatrix}
    A & B \\ C & D_{M0}(\Omega)
  \end{pmatrix}^{-1} &=
  \begin{pmatrix}
    A^{-1}+A^{-1}BS^{-1}CA^{-1} & -A^{-1}BS^{-1} \\ -S^{-1}CA^{-1} & S^{-1}
  \end{pmatrix}\label{inv1}  \\
 & =
  \begin{pmatrix}
    T^{-1} & -T^{-1}BD_{M0}^{-1} \\[2pt] -D_{M0}^{-1}CT^{-1} & D_{M0}^{-1}+D_{M0}^{-1}CT^{-1}BD^{-1}
  \end{pmatrix} \label{inv2}
\end{align}
by means of the Schur complements
\begin{align}
  S&=D_{M0}(\Omega)-C A^{-1} B=D_{M0}(\Omega)+\frac{\Gamma_M\kappa\Omega_M\Delta}{(\kappa-i\Omega)^2+\Delta^2}=:D_{M}(\Omega),\label{eq:Schur-S}\\
  T&=A-BD^{-1}_{M0}(\Omega)C=
  \begin{pmatrix}
    \kappa-i\Omega & \Delta \\
    -\Delta -\frac{\Gamma_M\kappa\Omega_M}{D_{M0}(\Omega)} & \kappa-i\Omega
  \end{pmatrix}.
\end{align}
The effective mechanical susceptibility including optically induced shift and broadening is
\begin{align}
\chi_M(\Omega)&=2\Omega_M   D_M^{-1}(\Omega).\label{eq:chi_M}
\end{align}

The intracavity quadratures following from Eqs.~\eqref{inv1} and \eqref{inv2} are

\begin{align*}
  {\bf \hat X}_L(\Omega)&=O_\phi T^{-1} O^T_\phi\left(\sqrt{2\kappa_1}{\bf \hat X}^M_{L,in}(\Omega)+\sqrt{2\kappa_2}{\bf \hat V}_{in}(\Omega)\right)
  +\frac{\sqrt{2\gamma_{M0}\kappa\Gamma_M}\Omega_m }{D_M(\Omega)}  O_\phi A^{-1} {\bf \hat F}(\Omega),
\end{align*}
in which
\begin{align*}
     {\bf \hat F}(\Omega)&:=\begin{pmatrix} 0 \\ \hat f(\Omega) \end{pmatrix},
\end{align*}
and the field reflected off the cavity in port $1$ is
\begin{align}
  {\bf \hat X}^M_{L,out}(\Omega)&=-{\bf \hat X}^M_{L,in}(\Omega)+\sqrt{2\kappa_1}{\bf \hat X}_L(\Omega)\nonumber\\
  &=O_\phi(2\kappa_1T^{-1}-\mathds{1})O^T_\phi{\bf \hat X}^M_{L,in}(\Omega)+\sqrt{4\kappa_1\kappa_2}O_\phi T^{-1}O^T_\phi{\bf \hat V}_{in}(\Omega) \nonumber \\ &\quad+\frac{2\sqrt{\Gamma_M\kappa\kappa_1\gamma_{M0}}\Omega_m}{D_M(\Omega)}  O_\phi A^{-1} {\bf \hat F}(\Omega)\nonumber\\
  &=:\mathds{M}(\Omega){\bf \hat X}^M_{L,in}(\Omega)+\mathds{V}(\Omega){\bf \hat V}_{in}(\Omega)
  +\mathds{F}(\Omega)  {\bf \hat F}(\Omega)\label{eq:inoutM}
\end{align}
 
  The optomechanical transfer matrix $\mathds{M}(\Omega)$ is explicitly given by
\begin{align}
      \mathds{M}(\Omega)&=      \frac{2\kappa_1}{(\kappa-i\Omega)^2+\Delta\left(\Delta+\frac{\Gamma_M\kappa\Omega_M}{D_{M0}(\Omega)}\right)}
      O_\phi\begin{pmatrix}
    \kappa-i\Omega & -\Delta \\
    \Delta +\frac{\Gamma_M\kappa\Omega_M}{D_{M0}(\Omega)} & \kappa-i\Omega
  \end{pmatrix}O^T_\phi-\mathds{1}\nonumber\\
  &=\frac{1}{D_c(\Omega)}
      \frac{2\kappa_1D_{M0}(\Omega)}{D_{M}(\Omega)}
        O_\phi\begin{pmatrix}
    \kappa-i\Omega & -\Delta \\
    \Delta +\frac{\Gamma_M\kappa\Omega_M}{D_{M0}(\Omega)} & \kappa-i\Omega
    \end{pmatrix}O^T_\phi-\mathds{1}\label{eq:OM-transfer}
\end{align}

    where $D_c(\Omega)=(\kappa-i\Omega)^2+\Delta^2$. In the form given in the second line the dependence on the effective mechanical susceptibility becomes evident.

    We note that for a broadband cavity ($\kappa\gg \Delta,\Omega_M,\Omega$) and neglecting losses ($\kappa_2=0$) one recovers from Eq.~\eqref{eq:inoutM} the simple optomechanical input-output relation stated in the main text,
    \begin{align}
    \begin{pmatrix} \hat X^M_{L,out}(\Omega) \\ \hat P^M_{L,out}(\Omega) \end{pmatrix} &=\begin{pmatrix} 1 & 0 \\ \Gamma_{M}\chi_{M}(\Omega) & 1 \end{pmatrix} \begin{pmatrix} \hat X^M_{L,in}(\Omega) \\ \hat P^M_{L,in}(\Omega) \end{pmatrix}+\sqrt{\Gamma_{M}\gamma_{M0}}\chi_{M}(\Omega)\begin{pmatrix} 0 \\ \hat f(\Omega) \end{pmatrix}.\label{eq:inoutMsimple}
    \end{align}
    In the limit considered here the susceptibility corresponds to the one of the bare mechanical system (without shift and broadening).

    For nonzero detuning and taking into account effects of a finite cavity linewidth, the more involved input-output relations described by Eqs.~\eqref{eq:inoutM} have to be considered in general. However, in the unresolved-sideband regime ($\kappa \gg \Omega_M,\Omega$) we may obtain a simplified expression for the optomechanical transfer matrix~\eqref{eq:OM-transfer}. To this end, we note that the cavity response to the individual sideband components at $\pm\Omega$ of the light quadratures~\eqref{eq:2phot-quad} is governed by the complex Lorentzian
\begin{align}
    L(\Omega) &= \frac{\kappa}{\kappa-i(\Delta+\Omega)}=:|L(\Omega)| e^{i\theta(\Omega)},&
  |L(\Omega)| &= \frac{\kappa}{\sqrt{\kappa^2+(\Delta+\Omega)^2}},&
  \theta(\Omega) &= \arctan \left( \frac{\Delta+\Omega}{\kappa} \right),
\end{align}
where we have introduced its polar decomposition. In terms of this, Eq.~\eqref{eq:OM-transfer} can be reexpressed as (again neglecting cavity losses for simplicity, $\kappa_{2}=0$)
\begin{multline}
\mathds{M}(\Omega)=e^{i[\theta(\Omega)-\theta(-\Omega)]}O_{\phi+[\theta(\Omega)+\theta(-\Omega)]/2}\Bigg([1+i\frac{\Gamma_M\chi_{M}(\Omega)}{4}(|L(\Omega)|^{2}-|L(-\Omega)|^{2})]\mathds{1}\\
+\frac{\Gamma_M\chi_{M}(\Omega)}{4}\left(\begin{array}{cc}
0 & -(|L(\Omega)|-|L(-\Omega)|)^{2}\\
(|L(\Omega)|+|L(-\Omega)|)^{2} & 0
\end{array}\right)\Bigg)O^T_{\phi-[\theta(\Omega)+\theta(-\Omega)]/2},\label{eq:M-polar}
\end{multline}
where $\phi = \theta(0)$ (see discussion of Eq.~\eqref{eq:phi} below) and $\chi_{M}(\Omega)$ is the effective mechanical susceptibility~\eqref{eq:chi_M}.
To obtain a simpler expression for Eq.~\eqref{eq:M-polar} in the regime $\kappa \gg \Omega_M,\Omega$, we expand $|L(\Omega)|$ and $\theta(\Omega)$ to linear order around the carrier frequency ($\Omega=0$ in the rotating frame),
\begin{gather}
|L(\Omega)|\approx L_{0}+\delta L(\Omega),\quad L_{0}:=|L(0)|=\frac{\kappa}{\sqrt{\kappa^{2}+\Delta^{2}}},\quad\delta L(\Omega):=-\frac{\Omega\Delta\kappa}{(\kappa^{2}+\Delta^{2})^{3/2}}\label{eq:L-amp-approx}\\
\theta(\Omega)\approx\phi+\delta\theta(\Omega),\quad\delta\theta(\Omega):=\frac{\Omega\kappa}{\kappa^{2}+\Delta^{2}},\label{eq:theta-approx}
\end{gather}
resulting in the optomechanical scattering matrix
\begin{equation}
\mathds{M}(\Omega)\approx e^{2i\delta\theta(\Omega)}O_{2\phi}\Bigg([1+i\Gamma_{M}\chi_{M}(\Omega)L_{0}\delta L(\Omega)]\mathds{1}
+\Gamma_{M}\chi_{M}(\Omega)L_{0}^{2}\left(\begin{array}{cc}
0 & 0\\
1 & 0
\end{array}\right)\Bigg),\label{eq:M-polar-NSB}
\end{equation}
to leading order in $\delta\theta,\delta L$ (the phase prefactor does not affect the resulting spectra and will be suppressed for brevity henceforth).

The transfer matrix in Eq.~\eqref{eq:M-polar-NSB} interpolates between the simple result in Eq.~\eqref{eq:inoutMsimple}, which is valid in the limit $\kappa\rightarrow\infty$, and the general result in Eq.~\eqref{eq:OM-transfer}.

When considering the field reflected off the cavity in port 1 in Eq.~\eqref{eq:inoutM}, the finite modematching $\eta_{mm}$ of the input quadratures ${\bf \hat X}_{L,in}(\Omega)$ to the cavity quadratures ${\bf \hat X}_{L}(\Omega)$ is treated as equivalent to the input port having higher loss, i.e. $\kappa_1 \rightarrow \eta_{mm} \kappa_1$. The total cavity loss remains fixed $\kappa = \kappa_1 + \kappa_2$ and we simply treat $\kappa_2 \rightarrow \kappa_2 + (1 - \eta_{mm})\kappa_1$ as the input for the additional vacuum noise. 

\subsection{Atomic Spins System}

As discussed in Ref.~\cite{Vasilakis2015,RMP}, in the limit of low saturation and large detuning from the atomic resonance, the Hamiltonian affecting the atomic spin and light polarization observables  can be written in the form:
\begin{equation}
\hat{H}_{\text{int}} =   \alpha \hat{S}_z \hat{J}_z, \label{eq:HamiltonianSpin0}
\end{equation}
where $\hat{J}_z$ is the dimensionless ($\hbar=1$) collective spin component along the direction of light propagation (taken here to coincide with the $z$ axis in the lab frame) and $\hat{S}_z$ is the Stokes component of light that measures the degree of circular polarization. The parameter $\alpha$ depends on the detuning from the resonance $\Delta_S$, on the area $A$ of interaction and on physical constants:
\begin{equation}
\alpha = \frac{\Gamma_{\text{sp}}}{8 A \Delta} \frac{\lambda^2}{2 \pi} \alpha_1(\Delta),
\end{equation}
where $\Gamma_{\text{sp}}$ is the spontaneous emission rate associated with the optical transition, $\lambda$ is the wavelength of light and $\alpha_1 (\Delta)$ is a numerical factor that depends on the specific atomic structure and for detunings much larger than the excited state hyperfine structure can be approximated to be unity.

For an ensemble of a large number of atoms, highly polarized along the direction of a static magnetic field ($x$ direction), the Holstein-Primakoff transformation can be performed and map the collective spin operators to position and momentum operators of an effective (spin) oscillator:
\begin{gather}
\hat{X}_S = \frac{\hat{J}_z}{\sqrt{|J_x|}}; \quad \hat{P}_S = -\text{sgn}(J_x)\frac{\hat{J}_y}{\sqrt{|J_x|}}. \label{eq:SpinOscillatorMapping}
\end{gather}
Here we represent the macroscopic mean polarization by its $x$-projection $J_x=\langle\hat{J}_x\rangle$ (including its sign) rather than merely its magnitude (as done in the main text for simplicity).
As described in the main text and above in the Methods section, the relative sign of $J_x$ and $B$ can be either positive or negative, reflecting whether the macroscopic spin is aligned or anti-aligned with respect to the applied magnetic field. For the case of this work, where Cesium atoms are polarized in the $F=4$ hyperfine manifold of the ground electronic state, positive $\text{sgn}(J_{x}/B)$ corresponds to a negative mass oscillator (energy should be extracted to remove the ensemble from the fully polarized state), whereas negative $\text{sgn}(J_{x}/B)$ corresponds to a positive mass oscillator.

The presence of the static magnetic field adds the Hamiltonian term
\begin{equation}\label{eq:eqLarmor}
    \hat{H}_S = \mu_B g_F  B \hat{J}_x 
\end{equation}
with $\mu_B$ being the Bohr magneton and $g_F=1/4$ the Land\'{e} factor for the $F=4$ manifold. In the language of spin oscillators this Hamiltonian term affects the evolution of $\hat{X}_S$ and $\hat{P}_S$ in the following way:
\begin{equation}
\dot{\hat{X}}_S\Big|_{B} =  \Omega_S \hat{P}_S; \quad \dot{\hat{P}}_S \Big|_{B} = -\Omega_S \hat{X}_S,
\end{equation}
where $\Omega_{S} = -\text{sgn}(J_{x}) \mu_B g_F  B$ so that $\Omega_{S}>0(<0)$ refers to the positive (negative) mass scenario.

A similar mapping can be performed with the Stokes components of light. For linearly polarized light in the $x$ direction with Stokes component $S_x=\text{sgn}(S_x) \Phi/2$, where $\Phi$ is the photon flux, the mapping can be written in the form:
\begin{equation}
\hat{X}_L = \frac{\hat{S}_z}{\sqrt{|S_x|}}; \quad \hat{P}_L = -\text{sgn}(S_x)\frac{\hat{S}_y}{\sqrt{|S_x|}}. \label{eq:LightOscillatorMapping}
\end{equation}
From Eqs.~\eqref{eq:HamiltonianSpin0}, \eqref{eq:SpinOscillatorMapping}, \eqref{eq:eqLarmor} \eqref{eq:LightOscillatorMapping} and \eqref{eq:ReadoutRate} we can write the Hamiltonian as
\begin{equation}
\hat{H} = \sqrt{\Gamma_S} \hat{X}_S \hat{X}_L + \Omega_S (\hat{X}^2_S + \hat{P}^2_S)/2,
\end{equation}
The input-output relationships for the Stokes components
\begin{equation}
\hat{S}_{z,\text{out}}(t) = \hat{S}_{z,\text{in}}(t); \quad \hat{S}_{y,\text{out}}(t) = \hat{S}_{y,\text{in}}(t)+ \alpha S_x \hat{J}_z(t) ,
\end{equation}
are mapped into:
\begin{equation}
\hat{X}_{L,\text{out}}(t) = \hat{X}_{L,\text{in}}(t); \quad \hat{P}_{L,\text{out}} = \hat{P}_{L,\text{in}}(t)+\sqrt{\Gamma_S} \hat{X}_S(t),\label{eq:spin-IO}
\end{equation}
where the readout rate $\Gamma_S$ is
\begin{equation}
\Gamma_S = \frac{1}{2} \alpha^2 \Phi |J_x| \label{eq:ReadoutRate}.
\end{equation}
The atomic spin dynamics, including the effects of its interaction with a Markovian reservoir, is
\begin{align}\label{eq:SpinDyn1}
    \dfrac{\mathrm{d}}{\mathrm{d}t}\hat{P}_S(t) &= -\Omega_S \hat{X}_S(t) - 2\gamma_S \hat{P}_S(t) + \sqrt{4\gamma_{S}}\hat{F}_{S}(t) + \sqrt{\Gamma_S} \hat{X}_L(t), \\
    \dfrac{\mathrm{d}}{\mathrm{d}t}\hat{X}_S(t) &= \Omega_S \hat{P}_S(t) \label{eq:SpinDyn2},
\end{align}
with $\hat{F}_{S}(t)$ being the random Langevin force acting on the spin; this force is the analogous of the thermal noise $\hat{f}$ that acts on the mechanical oscillator. Its correlation function is $\langle \hat{F}_S (t) \hat{F}_S (t') \rangle = \delta(t- t') (n_{S}+1/2)$, where a thermal spin occupancy $n_{S}>0$ reflects the excess noise induced by imperfect polarization of the ensemble.
In the above analysis, the effect of tensor polarizability in the evolution of the light and spin state was neglected. For the detuning used in the experiment ($\Delta_S \sim \si{3\giga\hertz}$) the effect of the tensor polarizability is estimated to be on the few percent level \cite{Marquardt2007}.

In frequency space, the spin system is structurally identical to the one of the simple limit considered in Eq.~\eqref{eq:inoutMsimple} for the optomechanical system, that is, Fourier transforming and solving Eqs.~(\ref{eq:spin-IO},\ref{eq:SpinDyn1},\ref{eq:SpinDyn2}), one obtains the matrix relationship
    \begin{align}\label{eq:eqMtrixSpins}
    {\bf X}^S_{L,out}(\Omega)&= \mathds{S}(\Omega){\bf X}_{L,in}(\Omega)+\sqrt{\Gamma_S\gamma_{S}}\chi_{S}(\Omega){\bf F}_S(\Omega). &
    \mathds{S}(\Omega)&=\begin{pmatrix} 1 & 0 \\ \Gamma_S\chi_{S}(\Omega) & 1 \end{pmatrix}
    \end{align}
    where the spin oscillator susceptibility is $\chi_{S}(\Omega)=2\Omega_S/(\Omega_S^2-\Omega^2-2i\Omega\gamma_S)$. The spin thermal noise is represented by ${\bf F}_S(\Omega) = [ 0,  \hat{F}_S (\Omega)]^T$. Here we adopt a phenomenological model for the susceptibility of the spin oscillator. A microscopic derivation along the lines of \cite{Vasilyev2012} would result in a slightly different susceptibility with corrections to the present one scaling as $Q_S^{-1}$ where $Q_S\gg 1$ is the quality factor of the atomic oscillator.
    
\subsection{Hybrid System}

     The two systems are connected such that ${\bf X}_{L,in}^M(\Omega)={\bf X}_{S,out}^M(\Omega)$, as shown schematically in Fig.~\ref{fig:TheorySchematic}. Taking into account losses and further phase shifts as indicated in the figure the compound transfer matrix for the hybrid optomechanical-spin system is
\begin{align}
&{\bf \hat X}_{L,out}(\Omega)=\nonumber\\
&\quad\sqrt{\eta_1\eta_2}\mathds{M}(\Omega)O_\varphi\mathds{S}(\Omega){\bf X}_{L,in}(\Omega) & &\text{(vacuum noise of light transduced through $S$ and $M$)}\nonumber\\
&\quad+\sqrt{\eta_1\eta_2}\mathds{M}(\Omega)O_\varphi \sqrt{\Gamma_S\gamma_{S}}\chi_{S}(\Omega){\bf F}_S(\Omega)& &\textrm{(spin noise transduced through $M$)}\nonumber\\
&\quad+\sqrt{(1-\eta_1)\eta_2}\mathds{M}(\Omega){\bf \hat V}_{1,in}(\Omega) & &\textrm{(vacuum noise of light from losses btw $S$ and $M$)}\nonumber\\
&\quad +\sqrt{\eta_2}\mathds{V}(\Omega){\bf \hat V}_{in}(\Omega) & &\textrm{(vacuum noise of light from losses in optomechanical cavity)}\nonumber\\
&\quad+\sqrt{\eta_2}\mathds{F}(\Omega)  {\bf F}(\Omega) & &\textrm{(thermal noise from $M$)}\nonumber\\
&\quad+\sqrt{1-\eta_2}{\bf \hat V}_{2,in}(\Omega) & &\textrm{(vacuum noise from losses between $M$ and detector)}\label{eq:hybrid-IO}
\end{align}
          where $\eta_1$ and $\eta_2$ denote the transmission efficiencies from the spin system to the optomechanical cavity and from the optomechanical cavity to the detector, respectively. Vacuum noises incurred through these losses are described by ${\bf \hat V}_{1(2),in}(\Omega)$. An optional phase shift $\varphi$ introduced deliberately in between the two systems is accounted for by the rotation matrix $O_\varphi$.

           Finally, the homodyne detection is performed in the frame of the classical field after the optomechanical system where it has acquired a phase shift relative to the field before the optomechanical cavity. This phase is found as follows: The classical intracavity amplitude $\alpha$ is connected to the incoming amplitude $\alpha_{in}$ by      
\begin{align}
\alpha&=\frac{\sqrt{2\kappa_1}}{\kappa-i\Delta}\alpha_{in}=\frac{\alpha_{in}}{\sqrt{\kappa^2+\Delta^2}}e^{i\phi}, & \phi&=\arctan(\Delta/\kappa).\label{eq:phi}
\end{align}
where $\kappa=\kappa_1+\kappa_2$. The outgoing field is
\[
\alpha_{out}=-\alpha_{in}+\sqrt{2\kappa_1}\alpha=\frac{\kappa_1-\kappa_2+i\Delta}{\kappa_1+\kappa_2-i\Delta}\alpha_{in}
=\frac{\left[\kappa_1-\kappa_2+i\Delta\right]\left[\kappa_1+\kappa_2+i\Delta\right]}{(\kappa_1+\kappa_2)^2+\Delta^2}\alpha_{in}\sim e^{i(\psi+\phi)}\alpha_{in}
\]
where $\psi=\arctan(\Delta/(\kappa_1-\kappa_2))$. Accordingly, the measured field quadrature $\hat{P}_{L,meas}$ is determined by
\[
\begin{pmatrix}
  \hat{X}_{L,meas} \\
  \hat{P}_{L,meas}
\end{pmatrix} = O_{\psi+\phi}^{T} {\bf \hat X}_{L,out}(\Omega).
\]
This relation is used to determine the measured noise spectral densities shown in the main text. For simplicity of notation, the measured quadrature $\hat{P}_{L,meas}$ is referred to as $\hat{P}_{L,\text{out}}$ in the main text and other parts of the Methods.

We will now use the transfer matrix of the hybrid system to analyse the QBA contribution to the optical output field [Eq.~\eqref{eq:hybrid-IO}, 1st line].
For the case when the opto-mechanical damping dominates the membrane response, $\gamma_M \gg \gamma_{M0}$, and in the sideband unresolved limit, $\Omega_M \ll \kappa$, we can apply the approximate optomechanical scattering matrix \eqref{eq:M-polar-NSB} to find (ignoring optical losses $\eta_1=0=\eta_2,\kappa_2=0$ for simplicity and setting $\varphi=0$)
\begin{equation}
\hat{P}_{L,meas} = \left[\Gamma_{M}L_0^2\chi_{M}(\Omega)+\Gamma_{S}\chi_{S}(\Omega)\left\{1 + i \Gamma_M \chi_M(\Omega) L_0 \delta L\right\}\right]\hat{{X}}_{L,in},\label{eq:BA-out}
\end{equation}
where $L_0$ is the empty cavity Lorentzian response and $\delta L$ is the difference in cavity response at frequencies $\pm\Omega\ll\kappa$. Only with $\delta L=0$ (LO$_2$ tuned to cavity resonance) the spin QBA and the mechanical QBA add/subtract in $\hat{P}_{L,meas}$. From Eqs.~(\ref{eq:D_M0},\ref{eq:Schur-S},\ref{eq:chi_M}) one finds that $L_0 \delta L =(\gamma_ M - \gamma_ {M0})/ \Gamma_M$, that is the distortion of the QBA due to $\delta L \neq 0$ has the same origin as the optomechanical broadening. In the relevant case of strong optomechanical cooling $\gamma_M \gg \gamma_{M0}$, there is no back action cancellation at the exact joint resonance frequency since $i\chi_{M}(\Omega=\Omega_M)\gamma_M=-1$.
In this regime the QBA power spectum of the hybrid system $S_{P_{L,meas}}$ corresponding to Eq.~\eqref{eq:BA-out} becomes
\begin{equation}
S_{P_{L,meas}} = \frac{(\Gamma_M \delta_S+\Gamma_S \delta_M)^2+\Gamma_M^2 \gamma_S^2  }{(\delta_{M}^{2}+\gamma_M^2)(\delta_S^2+\gamma_S^2)} S_{X_{L,in}},
\end{equation}
with $\delta_{M,S}=\Omega-\Omega_{M,S}$ and $S_{X_{L,in}}$ being the power spectral density of the input light amplitude fluctuations.
For matched responses ($\Gamma_S = \Gamma_M L_{0}^2$)
, $\Omega_M=\Omega_{S}$, $\Gamma_M=\Gamma_{S}, \gamma_M=\gamma_{S}$,
the ratio of the hybrid QBA spectrum to the QBA spectrum of the mechanics becomes $\gamma_M^2/((\Omega-\Omega_M)^2+\gamma_M^2)$, thus QBA evasion is indeed expected everywhere, except for $ \Omega=\Omega_M$. The minimal variance of the hybrid QBA is $1/2$ of the QBA of the mechanical oscillator alone.

\subsection{Calibration of quantum back action for the spin system}\label{sec:calibrationspins}

To characterize the quantum cooperativity, $C_q^S$, and the readout rate, $\Gamma_S$, one can use the fact that a single light quadrature is coupled to the oscillator: by suitable modulation of $\hat{X}_{L,\text{in}}$ one can boost the contribution of the measurement-induced back action.

As thoroughly discussed in the previous sections and summarized by equation \eqref{eq:eqMtrixSpins}, the input-output relations for the continuous readout of a harmonic oscillator are
\begin{align}\label{eq:masterEq}
    \left(\begin{array}{c}
X_{L,\text{out}} \\
P_{L,\text{out}}
\end{array}\right) &= \left(\begin{array}{c}
X_{L,\text{in}} \\
P_{L,\text{in}}
\end{array}\right) + \Gamma v_1^T L v_2 \left(\begin{array}{cc}
0 & 0 \\
1 & 0
\end{array}\right) \left(\begin{array}{c}
X_{L,\text{in}} \\
P_{L,\text{in}}
\end{array}\right) + \sqrt{\Gamma \gamma} v_1^T L F_{Th} \left(\begin{array}{c}
0  \\
1 
\end{array}\right),
\end{align}
in which $\Gamma$ is the readout rate, $\gamma$ the decay rate and
\begin{align*}
    L &= (i\omega 1 - M)^{-1} \\
    M &= \left(\begin{array}{cc}
0 & \omega_0 \\
-\omega_0 & -\gamma
\end{array}\right) \\
    v_1 &= \left(\begin{array}{c}
1 \\
0
\end{array}\right) \hspace{1cm} v_2 = \left(\begin{array}{c}
0 \\
1
\end{array}\right).
\end{align*}
In a more straightforward language, equation \eqref{eq:masterEq} becomes
\begin{align}\label{eq:eq0}
\left(\begin{array}{c}
X_{L,\text{out}} \\
P_{L,\text{out}}
\end{array}\right) &= \left(\begin{array}{c}
X_{L,\text{in}} \\
P_{L,\text{in}}
\end{array}\right) + R_\text{BA} \left(\begin{array}{cc}
0 & 0 \\
1 & 0
\end{array}\right)  \left(\begin{array}{c}
X_{L,\text{in}} \\
P_{L,\text{in}}
\end{array}\right) + R_\text{Th} \left(\begin{array}{c}
0  \\
1 
\end{array}\right),
\end{align}
with $R_\text{BA}$ and $R_\text{Th}$ being the response functions of the oscillator to the back action and thermal forces. 

The effect of losses is also important, as there is an admixture of uncorrelated vacuum, indicated by the subscript $v$, with the signal of interest; therefore
\begin{align}\label{eq:eq1}
\left(\begin{array}{c}
X_{L,\text{out}} \\
P_{L,\text{out}}
\end{array}\right) &\rightarrow \sqrt{\eta} \left(\begin{array}{c}
X_{L,\text{out}} \\
P_{L,\text{out}}
\end{array}\right) + \sqrt{1-\eta} \left(\begin{array}{c}
X_{L,\text{v}} \\
P_{L,\text{v}}
\end{array}\right).
\end{align}
The PSD for both light quadratures are calculated from the absolute square of the equation \eqref{eq:eq0}:
\begin{align}
    S_{XX} &= \eta \langle X_{L,\text{in}} X_{L,\text{in}}^\dagger \rangle + (1-\eta) \langle X_{L,\text{v}} X_{L,\text{v}}^\dagger \rangle \nonumber \\
    S_{PP} &= \eta \left[\langle P_{L,\text{in}} P_{L,\text{in}}^\dagger \rangle + R_\text{BA}^2 \langle X_{L,\text{in}} X_{L,\text{in}}^\dagger \rangle + R_\text{Th}^2 \right] + (1-\eta)\langle P_{L,\text{v}} P_{L,\text{v}}^\dagger \rangle.
    \label{eq:eq2}
\end{align}
Therefore, it is explicit that to boost the back action component of the oscillator readout in comparison to the other noise contributions, one should modulate the in-phase quadrature of light, $\hat{X}_L$. Doing so, the input spectral densities are
\begin{equation*}
    \langle X_{L,\text{in}} X_{L,\text{in}}^\dagger \rangle \rightarrow (n_{WN} +1) \langle X_{L,\text{in}} X_{L,\text{in}}^\dagger \rangle,
\end{equation*}
and $\langle X_{L,i}(\omega) X_{L,j}^\dagger(-\omega') \rangle = \delta(\omega-\omega')\delta_{ij}$, in which $i,j$ represent the different sources of fluctuations. Therefore, the input-output relations from \eqref{eq:eq2} are
\begin{align}
    S_{XX} &= \eta n_{WN} +1 \nonumber \\
    S_{PP} &= \eta \left[ R_\text{BA}^2 (n_{WN}+1) + R_\text{Th}^2 \right] + 1.
    \label{eq:eq3}
\end{align}
Experimentally, to be able to calculate the back action to thermal noise ratio, one needs to measure the (i) response of the system to SN drive and (ii) the response of the system with some known modulation $n_{WN}$, for a given probe power. Calibrating the curves in shot noise units, the measured spectral on-resonance heights after the subtraction of the white noise contribution, defined here as A and B, are
\begin{align*}
    S_{PP}^{WN} -1 &= B= \eta \left[ R_\text{BA}^2 (n_{WN}+1) + R_\text{Th}^2     \right] \\
    S_{PP}^{SN} -1 &= A= \eta \left[ R_\text{BA}^2  + R_\text{Th}^2     \right],
\end{align*}
therefore, the desired ratio is:
\begin{align}
    \dfrac{R_\text{BA}^2}{R_\text{Th}^2} = \dfrac{B-A}{(n_{WN}+1)A-B}.
\end{align}

This technique was used to calibrate $C_q^S$, the quantum cooperativity of the spin oscillator. For this measurement, $\eta=0.7$ is the detection efficiency, $S_{\text{WN}}$ is the spectral density of added white noise in units of vacuum noise and $S_{QBA}$ and $S_{\text{TH}}$ are the back action and thermal spectral densities, respectively. The measurements of the phase noise presented in (Figure~2c) and of $S_{\text{WN}}$ are performed with polarization interferometry using LO$_1$ calibrated to the shot noise of LO$_1$.  From the phase noise $S_{\text{PP},0}$ and $S_{\text{PP},N}$ measured for $n_{\text{WN}}=0$ (vacuum input) and $S_{\text{WN}}=1.2$, respectively (Figure~2c), we find $S_{\text{QBA}}=(S_{\text{PP},N}-S_{\text{PP},0})/(1.2 \times \eta_A
)$ and $C_q^S =(S_{\text{PP},N}-S_{\text{PP},0})/(2.2\times S_{\text{PP},N}-S_{\text{PP},0})=1.10\pm 0.15$. 

\bibliographystyle{naturemag}
% \bibliography{references}

% \setcounter{page}{1}
% {\centering\textbf{Extended Data\label{sec:ExtendedData}}}
% \section*{Extended Data}\label{sec:ExtendedData}

\newpage

\begin{table}
    \centering
\begin{tabular}{SSS}  
    {\textbf{Parameter}} & {\textbf{\text{Symbol}}} & {\textbf{Value (Fig. 2\&3, Fig. 4)}} \\ \midrule
    & {\textbf{Atomic spin oscillator}} &  \\
%    \bottomrule

    {Intrinsic decoherence rate}  & {$\gamma_{S0}$} & {$2\pi \times 500$\si{\hertz}}  \\
    {Total decoherence rate}  & {$\gamma_{S}$} & {$2\pi \times (2.6, 2.3)$\si{\kilo\hertz}} \\
    {LO$_1$ driving power}  &   &  {$(1.7, 1.5)$\si{mW}} \\
    {Detuning from the $D_2$ $F=4 \rightarrow F'=5$ transition}  & $\Delta_S$  & {3\si{GHz}}  \\
    {Quantum cooperativity}    & {C$_q^S$}  & {$1.10\pm 0.15$}  \\
    {Spin Polarization}    &   &  {60\%} \\
    {Microcell optical losses}   & {$\eta_{\text{microcell}}$}  &  {13\%} \\
    {Microcell temperature}    &   &  {65\si{\degreeCelsius}}  \\
    \midrule
    
    & {\textbf{Mechanical oscillator}} & \\
%    \bottomrule
    {Effective mass} & {$m_{\text{eff}}$} & {$14\si{ng}$}\\
    {Zero point fluctuations} & {$x_{\text{zpf}}$} & {$1\si{fm}$}\\
    {Intrinsic mechanical frequency}  & {$\Omega_{M}$} & {$2\pi \times 1.28\si{MHz}$} \\
    {Intrinsic damping rate}  & {$\gamma_{M0}$} & {$2\pi \times 50\si{mHz}$} \\
    {Cavity detuning}  & $\Delta$  & {$2\pi \times -4.7\si{MHz}$}  \\
    {Total cavity half linewidth}  & {$\kappa$} & {$2\pi \times (8.7, 7.7)\si{MHz}$} \\
    {LO2 drive power}  & { } & {$(54, 38)\si{\micro \watt}$} \\
    {Intracavity photons} & $N$ & {$(5.7, 4.2)\times 10^6$}\\
    {Single photon coupling rate}  & {$g_0$} & {$2\pi \times 210\si{Hz}$} \\
    {Thermal bath temperature} & {$T_\text{bath}$} & {$7\si{K}$} \\
    {Bath occupancy}  & {$n_\text{bath}$}  & {$114\times 10^3$} \\
    {Quantum cooperativity} & {C$_q^M$}  &  {(2.6, 2.2)} \\

%    {Readout rate} & {$\Gamma_M$} &  \\
    {Mechanical linewidth} & {$\gamma_M$} & {$2\pi \times 2.7\si{kHz}$}\\
    {Mean thermal occupancy}  & {$\bar{n}_M^{th}$} & {2.1} \\

    \midrule
    
    & {\textbf{Hybrid \& detection}} &  \\
%    \bottomrule
    {Quantum efficiency in between systems}  &   & {$61\%$} \\
    {Detection efficiency}  &  & {$72\%$, $75\%$} \\
    {Homodyning visibility}  &  & {$89\%$}  \\
    {Cavity mode-matching} & {$\eta_{mm}$} & {$90\%$} \\
    \bottomrule
\end{tabular}
\caption{Summary of notation and experimental parameters.}
\label{table:parameters}
\end{table}